\documentclass{aastex}          %% The manuscript based on AASTeX v5.x
\usepackage{spr-astr-addons}    %% mimicing ASTR journal style
\newcommand{\zf}{z$_{\rm{F}}$} % redshift at the foreground quasar
\newcommand{\zb}{z$_{\rm{B}}$} % redshift at the background quasar
\newcommand{\qsof}{QSO$_{\rm{F} }$} % foreground quasar
\newcommand{\qsob}{QSO$_{\rm{B}}$} % background quasar
\newcommand{\mgII}{MgII }

\newcommand{\ewrest}{EW$_{\rm{rest}}$}
\newcommand{\ewrestI}{EW$_{\rm{rest}}$($\lambda$2796)}
\newcommand{\ewrestII}{EW$_{\rm{rest}}$($\lambda$2803)}
\newcommand{\zabs}{z$_{\rm{abs}}$}
\newcommand{\lamI}{$\lambda_{abs}(\lambda2796)$}
\newcommand{\lamII}{$\lambda_{abs}(\lambda2803)$}
\newcommand{\FFp}{(Farina et al. 2013, 2014)}
\newcommand{\FF}{Farina et al. (2013, 2014)}

\begin{document}
%% Article title
%
\title{The circum-galactic medium of quasars: transverse and line-of-sight absorptions}

%% Running heads
\shorttitle{The circum-galactic medium of quasars: transverse and line-of-sight absorptions}
\shortauthors{<Autors et al.>}

%% Author and Affilations
\author{A. Sandrinelli\altaffilmark{1}}
 \and \author{R. Falomo\altaffilmark{2}}
 \and \author{A. Treves\altaffilmark{1,3}}
 \and \author{S. Paiano\altaffilmark{3}} 
 \and \author{R. Scarpa\altaffilmark{4,5}}  %\and 
%\author{\altaffilmark{}}
%\affil{}
\email{asandrinelli@yahoo.it} %% non-output

%% Alternate Affilations
\altaffiltext{1}{INAF, Istituto Nazionale di Astrofisica -- Osservatorio Astronomico di Brera, via E. Bianchi 46, I-23807 Merate, Italy}
\altaffiltext{2}{INAF, Istituto Nazionale di Astrofisica -- Osservatorio Astronomico di Padova, Vicolo dell'Osservatorio 5, I-35122 Padova (PD), Italy}
\altaffiltext{3}{Dipartimento di Scienza e Alta Tecnologia (DISAT), Universit\`a degli Studi dell'Insubria, via Valleggio 11,I-22100 Como, Italy}
\altaffiltext{4}{Instituto de Astrofısica de Canarias, c/via Lactea s/n
San Cristobal de la Laguna, 38205, Spain}
\altaffiltext{5}{Departamento de Astrofisica, Universidad de La Laguna (ULL), 38206 La Laguna, Tenerife, Spain}

%% Abstract
\begin{abstract}

Quasar projected pairs (QPPs) can be used for investigating the circumgalactic medium of  quasars through the study of intervening 
absorption lines in the spectrum of the background quasar (\qsob) that are at the same redshift of the foreground quasar (\qsof).
 Here we report on optical spectroscopy, gathered at Gran Telescopio Canarias,  of 14 QPPs.
  In 7 cases we find \mgII absorption lines associated with the foreground quasar. Only for two cases line-of-sight absorptions (LOS) are revealed. 
These new observations complement our previous study performed on other 30 QPPs.
A brief discussion of the properties of the intervening absorption lines associated with the foreground quasar for the full dataset is reported.

%with the foreground quasar at z $\sim$ 1 and observed in a region where \mgII of the \qsof \ %is apparent, enlarging the sample 
%of 28 QPPs presented in previous papers of ours. 
%In 8  cases we find the transverse absorptions, while in only two cases line-of-sight %absorptions (LOS) are revealed. 

\end{abstract}

%% Keywords

\begin{table*}
\scriptsize
%\tiny
\caption{The sample of quasar projected pairs and properties.
In columns: identification label of the system (ID), position of the foreground target, redshift derived from literature (z), V-band apparent and absolute magnitude, angular ($\Delta \theta$) and projected (R$_\bot$) 
separation between the two quasars, seeing during the observations (see), signal-to-noise ratio  (SN) of the spectrum. The labels F and B refer to the foreground
and background quasar, respectively.
\label{sample}}
\begin{tabular}{@{}ccllcccccccll@{}}
\tableline
\tableline
    ID  &\qsof \ coordinates    & \zf $$& \zb$$  &  V$\rm{_F}$  &V$\rm{_B}$& M(V)$\rm{_F}$  &M(V)$\rm{_B}$& $\Delta\theta$ &R$_\bot$      &    see &SN$\rm{_F}$&SN$\rm{_B}$    \\
   	&(J2000)                                             &             &        &  (mag)          &(mag)         &  (mag)                &(mag)             &(arcsec) & (kpc)      &(arcsec)  &         & \\
\hline
QQ01   & 08:36:49.47$+$48:41:50.1   &  0.657      &   1.711  &  19.14 &  18.66 &   -23.71 & -26.59	&  3.7  &  25     &   0.8  & 20  & 30  \\  
QQ02   & 08:59:15.10$+$42:41:23.6   &  0.901       &   1.396    &  19.37 &  20.79 		 & -24.18 & -23.96	 &  3.6  &  30     &   0.8  & 30   & 10 \\ 
QQ03   & 09:28:27.92$-$00:11:26.5 &  0.878         &   1.138  &  19.24 &  20.24  	& -24.16 & -23.91	&  28.6 &  221    &   1.6  & 10  & 10 \\    
QQ04   & 10:19:57.47$-$02:43:05.5 &  0.863         &   0.993    &  20.58 &  20.20  		& -23.29 & -23.61	&  25.4 &  196    &   1.3  & 10  & 15 \\ 
QQ05   & 10:27:53.84$+$00:30:55.4   &  1.129       &   1.904   &  18.63 &  21.30 	 &-25.50  &-24.77	 &  28.9 &  237    &   1.9  & 95  & 10 \\ 
QQ06  & 10:44:31.74$+$61:38:48.4   &  1.096       &   2.522  &  20.07 &  20.38  	& -23.98 & -25.99	&  13.7 &  112    &   0.8  & 15  & 15  \\
QQ07 & 11:14:43.15$ -$00:51:21.5  &  0.835         &   1.815    &  20.26 &  20.70  		& -23.52 & -25.24	&  27.3 &  208    &   1.3  & 10  & 20 \\
QQ08 & 13:33:16.10$+$00:36:24.9   &  1.030        &   1.794   &  20.98 &  20.58		 &-22.92  &-24.80	 &  16.9 &  136    &   2.5  & 6  & 10 \\
QQ09 & 13:39:45.07$+$00:10:04.5   &  0.978         &   1.873 &  20.02 &  19.23 		& -23.75 & -26.29	&  19.0 &  151    &   2.4  & 10  & 23  \\ 
QQ10 & 15:45:43.05$+$05:49:29.0   &  0.644         &   1.226  &  19.71 &  18.99  		& -23.09 & -25.37	&  28.7 &  198    &   1.8  & 20  & 30   \\  
QQ11& 17:11:29.32$+$29:15:23.0  &  1.117         &   2.163  &  20.32 &  20.52  		& -23.78 & -25.42	&  27.9 &  229    &   0.8  & 15  & 10  \\  
QQ12 & 21:20:27.06$-$00:19:51.1  &  0.86          &   2.578  &  20.69 &  17.77  		& -22.76 & -28.67	&  19.3 &  148    &   0.7  & 10  & 45 \\ 
QQ13 & 23:12:52.80$+$14:44:58.6   &  0.768       &   1.523  &  19.93 &  17.74 	& -23.28 & -27.24	&  6.3  &  47     &   0.8  & 10   & 105 \\
QQ14  & 23:13:03.84$+$10:49:15.5   &  0.713       &   1.333    &  19.70  &  18.90  	& -23.35 & -25.71	&  11.3 &  82     &   1.8  & 20 & 20 \\  
\tableline
\end{tabular}
\end{table*}

\begin{table*}
\centering
\small
\caption{Properties of  \mgII transverse absorption features in the 
spectrum of the background  QSO and associated to the foreground QSO.
 The columns show the observed wavelengths ($\lambda$$_{\mathrm{abs}}$), 
 the rest-frame equivalent width  (\ewrest),  
 the redshift of the absorber (\zabs). 
The 2$\sigma$ upper limit for the minimum equivalent width  \ewrest \ detectable
   in the background QSO spectrum is also quoted. 
\label{abs}}
\begin{tabular}{@{}lcccccc@{}}
\tableline
\tableline
   ID   & \lamI   &\ewrestI 		& \lamII & \ewrestII   		& \zabs	   &EW$_{\rm{min,rest}}$ \\
   	&(\AA)	  & (\AA)   		& (\AA)  & (\AA)       		&      	   &(\AA)\\
\hline
QQ01    &  4629.8 & 1.84 $\pm$  0.10 	& 4641.0 & 1.56 $\pm$ 0.08 	&0.6556   &   0.13 \\ %  J0836
QQ02    &  5326.0 & 1.17 $\pm$  0.18   	& 5339.3 & 1.57 $\pm$ 0.39   	&0.9046   &   0.38 \\ %  J0859
QQ03    &  -	  & -			&-       & -	 		& -	   &   0.27 \\ %  J0928
QQ04    &  -	  & -			&-    	 & -	 		& -	   &   0.42 \\ %  J1019
QQ05    &  -	  & -			&-    	 & -	 		& -	   &   0.36 \\ %  J1027
QQ06    & -       & -			& -	 & -     		& -	   &   0.19 \\ %  J1044
QQ07    &  5126.6 & 0.51 $\pm$ 0.24 	& 5141.6 & 0.44 $\pm$ 0.19 	&0.8340   &   0.19 \\ %  J1114
QQ08    &  5667.0 & 1.64 $\pm$ 0.15	& 5679.8 & 1.18 $\pm$ 0.21  	& 1.0263  &   0.76 \\ %  J1333
QQ09    & -  	  & -			& -	 & -  			& -	   &   0.20 \\ %  J1339
QQ10    &  -      & - 	                & -      &- 			& -	   &   0.19 \\ %  J1545 
QQ11    &  5924.4 & 0.51 $\pm$ 0.15 	& 5940.7 & 0.51 $\pm$ 0.27	&  1.1190 &   0.49 \\ %  J1711
QQ12    &  -	  & -			& -      & -	 		& -	   &   0.11 \\ %  J2120
QQ13    &  4940.0 & 0.36 $\pm$ 0.05  	& 4952.7 & 0.21 $\pm$ 0.06 	& 0.7666  &   0.06 \\ %  J2312
QQ14    &  4788.4 & 0.71 $\pm$ 0.16 	& 4801.2 & 0.44 $\pm$ 0.18  	& 0.7125  &   0.29 \\ % J2313 
 \tableline
\end{tabular}
\end{table*}

\begin{table*}
\centering
\small
\caption{Properties of \mgII line-of-sight absorption systems  in the foreground QSOs.
  The radial velocity difference  $\Delta$V between the absorption-system and the foreground
QSO is added. 
\label{los}}
\begin{tabular}{@{}lccccccc@{}}
\tableline
\tableline
ID   & \lamI  &\ewrestI 	& \lamII  & \ewrestII   	& \zabs		& $\Delta$V	&EW$_{\rm{min,rest}}$   \\
     &		(\AA)	        & (\AA)   & (\AA)                          	&(km/s) 	&(\AA)			\\
\hline
QQ07 & 5128.6 & 0.43 $\pm$ 0.11 & 5142.2  & 0.25 $\pm$ 0.15 	& 0.8341 	&  -250 	&   0.35		\\ % J1114
QQ09 & 5493.9 &	0.35 $\pm$ 0.10	& 5508.3  & 0.46 $\pm$ 0.14 	& 0.9647	& -2000		&   0.32 		\\ % J1339 \hline
 \tableline
\end{tabular}
\end{table*}

\section{Introduction}\label{theintro}
  
 The circum-galactic medium (CGM) is the diffuse gas spanning a few hundred kpc  
 outside the galaxies and represents the  most abundant reservoir of baryons of galaxies \citep[e.g.][]{Peek2015,Telford2019}.
It is the link between the interstellar  and the intergalactic media,  so that  gas 
exchanges between them  have to go through it and necessarily leave here 
the proof of its passage \citep[see e.g.][for a review]{Tumlinson2017}. 
These exchanges  play an important role in the evolution of a galaxy because 
they can favour or suppress different processes in the galaxy, notably star formation.
Various  feedback mechanisms are invoked in hydrodynamic simulations
 for discriminating between possible theoretical  models \citep[e.g.][]{Oppenheimer2010,Fielding2017}.

 An effective  method for probing the  properties of the CGM of a galaxy  is to study the  absorption spectrum of a background 
 source which is angularly close so that the projected distance (R$_\bot$)  is of the order of hundreds kpc.
Quasars are sources apt to this scope since they are bright and point-like. 
 If one is interested in the CGM of a quasar, the system is a quasar projected pair (QPP): 
 the background quasar (\qsob)   is the light source, which enables to explore the GCM of the foreground quasar (\qsof) 
 at different projected distance, or impact parameter, from the \qsof. 
 This approach was introduced by \citet{Hennawi2006}, who studied the  
Ly$\alpha$ absorption systems in the halo of quasars at z$\sim$2
and found a high fraction  of absorbers for small separations (R$_{\bot}<$150 h$^{-1}$ kpc)
 coincident with the foreground quasar. This 
 provided significant evidence that these absorbers are strongly clustered around quasars.  
Using  10m class telescopes we investigated the presence of intervening \mgII absorption systems in the CGM of QSOs  
at  z$\sim$1 \citep{Farina2013,Farina2014}  by examining 30  
QPPs with angular separations  10$\arcsec$ - 30$\arcsec$ in order to explore the CGM in the 
 50-200 kpc region.  In about half cases \mgII  absorptions due to the foreground quasar
 were detected in the background quasar spectrum (traverse absorptions). 
 The  indication is that the covering  of the gas drops quickly beyond 100 kpc. In only one case \mgII absorption 
 along the line-of-sight (LOS) was revealed.
These results were also confirmed by \cite{Johnson2015} who carried out a study of a data set of QPPs from SDSS DR12 
spectral archives.  They  found a strong dependence of the absorption equivalent width on the luminosity/mass 
of the \qsof, point which was already suggested in \cite{Farina2014}.

In this paper we report observations of 14 new QPPs obtained at the  10.4m Gran Telescopio Canarias (GTC) (Sect. \ref{thesample}).
Our  main aim is to enlarge our previous  sample and in  compare the  transverse and LOS \mgII absorption lines. 
For this work we assume the following cosmology:  
$\rm{H}_{0}=~70~\rm{km}~\rm{s}^{-1}~\rm{Mpc}^{-1}$, 
\  $\Omega_{\rm{m}}=0.3$ and $\Omega_{\Lambda}=0.7$.

\section{Properties of the projected quasar pairs}\label{thesample}

In order to extend the data set of our previous work  \FFp \ 
towards small (R$_{\bot}\lesssim$ 50  kpc) and large (R$_{\bot}\gtrsim$ 200 kpc) projected distances, 
we searched for  new targets in The Million Quasars  Catalog \citep{Flesch2015}, following the selection criteria given below.
We required projected distance of the pairs to be  less than 250 kpc 
at the redshift of foreground quasar, which allows us to
explore the outer regions of a  CGM halo. 
 Differences between the radial velocities of the two  sources  were chosen
 $ >$ 5000 km/s to ensure the quasar pair is not a gravitational bound system.
The redshifts are drawn from SDSS archives, when present, otherwise from literature.

\begin{figure}[h]
\centering
%\vspace{2.9cm}
\includegraphics[width=1\columnwidth]{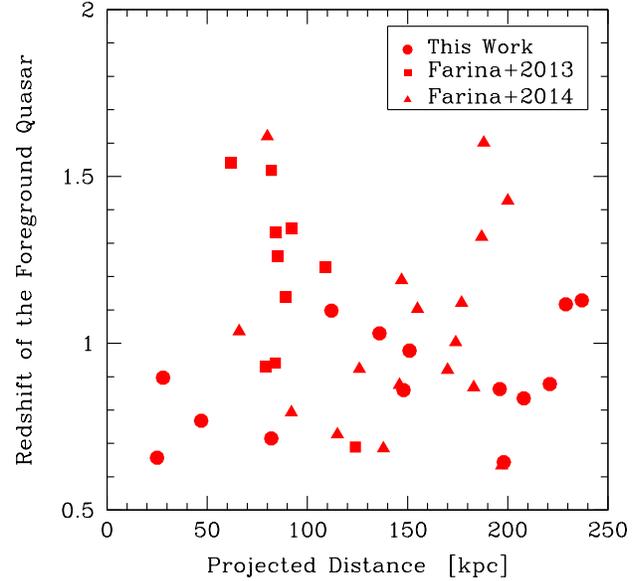}
%\vspace{-2.9cm}
\caption{
 Distribution in the R$_{\bot}$–z$_F$ plane of the projected QSO pairs 
studied to investigate the MgII absorbing circumgalactic medium of quasars. Triangles and squares are objects  investigated in  \cite{Farina2013}
and  \cite{Farina2014}, respectively.
Newly observed targets from this work are reported as circles.
}
\label{pdz}
\end{figure}

 The  \mgII doublet absorption systems  in the background quasar spectrum, 
associated to the  \mgII emission in the foreground quasar, 
have to fall inside the spectral range (4500-6000 \AA)  of the adopted grism (see Subsect. \ref{theobs}) 
 constraining   the foreground quasars at redshift interval 0.6$<$z$<$1.15. 
Paired targets are also selected ensuring 
good visibility from the GTC site in Roque de los Muchachos, La Palma, 
and  request that the background quasar be brighter than V=20.8 
in order to obtain spectra with signal-to-noise ratio $\gtrsim$ 10 in normal weather conditions.
From this new selection  14 QQPs have been targeted at the GTC, yielding the full data set  given in Fig.\ref{pdz} in the R$_{\bot}$-z plane.
In Table \ref{sample} we report the  general properties of the newly observed pairs.

% FIGURE ------

\section{Observations and data reduction}\label{theobs}

The QPP  spectra were secured at the GTC 
from Sept. 2017 to Aug. 2018 equipped with the Optical System
for Imaging and Low Resolution Integrated Spectroscopy  \citep[OSIRIS,][]{Cepa2003} R2500V grism, 
yielding effective intermediate spectral resolution R = 1500 for a slit  of 1.00", with 1 px corresponding 
to 1.3 \AA\ or to $\sim$ 70 km s$^{-1}$ at the central spectral wavelength.
This allows us to fully resolve the \mgII doublet components  
($\lambda$2797 \AA \ and $\lambda$2803 \AA). The spectral window covers the 4500-6000 \AA \ range.
 For each QPP the slit was oriented to simultaneously gather the spectrum of both objects.
 The full observation of each QSO pair was divided into three exposures  (3 $\times$ 1200 s)
  in order to provide an optimal correction of cosmic rays and CCD defects. 
The seeing obtained were $\sim$ 1.5 (see Table \ref{sample}).
We reduced  data by using the  standard IRAF recipes, using CCDRED package for bias
 subtraction and flat field correction.
 Wavelength calibration has been performed through 
 Xe+Ne+HgAr arc lamps and 
flux calibration was assessed using standard stars observed
 during the same nights of the targets. Corrections for systematics, slit losses
and variation of the sky conditions have been introduced  through aperture photometry
of the field  acquired concurrently to the spectra.

The GTC spectra, normalised to the continuum level, are showed in Fig. \ref{spec}.

\section{Transverse absorption systems}\label{theabs}

We search for MgII($\lambda2797$, $\lambda2803$)
absorption features with  \ewrestI \ larger than the minimum observable equivalent width
  \citep[EW$_{min}$, see][]{Paiano2017} in the \qsob \ spectrum.

\begin{figure*}
%\vspace{-1cm}
\centering
\caption{
Close-ups of the normalized \qsob spectra (red lines) presenting transverse absorption systems  associated to the 
\mgII  emission line of the \qsof. 
Gaussian fits are  performed on the absorptions lines and drawn as solid black lines. Dotted black lines 
indicate the positions of  spectral absorption peaks, while the green dashed line marks the position of  the \mgII emission line
in the  \qsof \ spectrum.
}
\includegraphics[width=0.44\columnwidth]{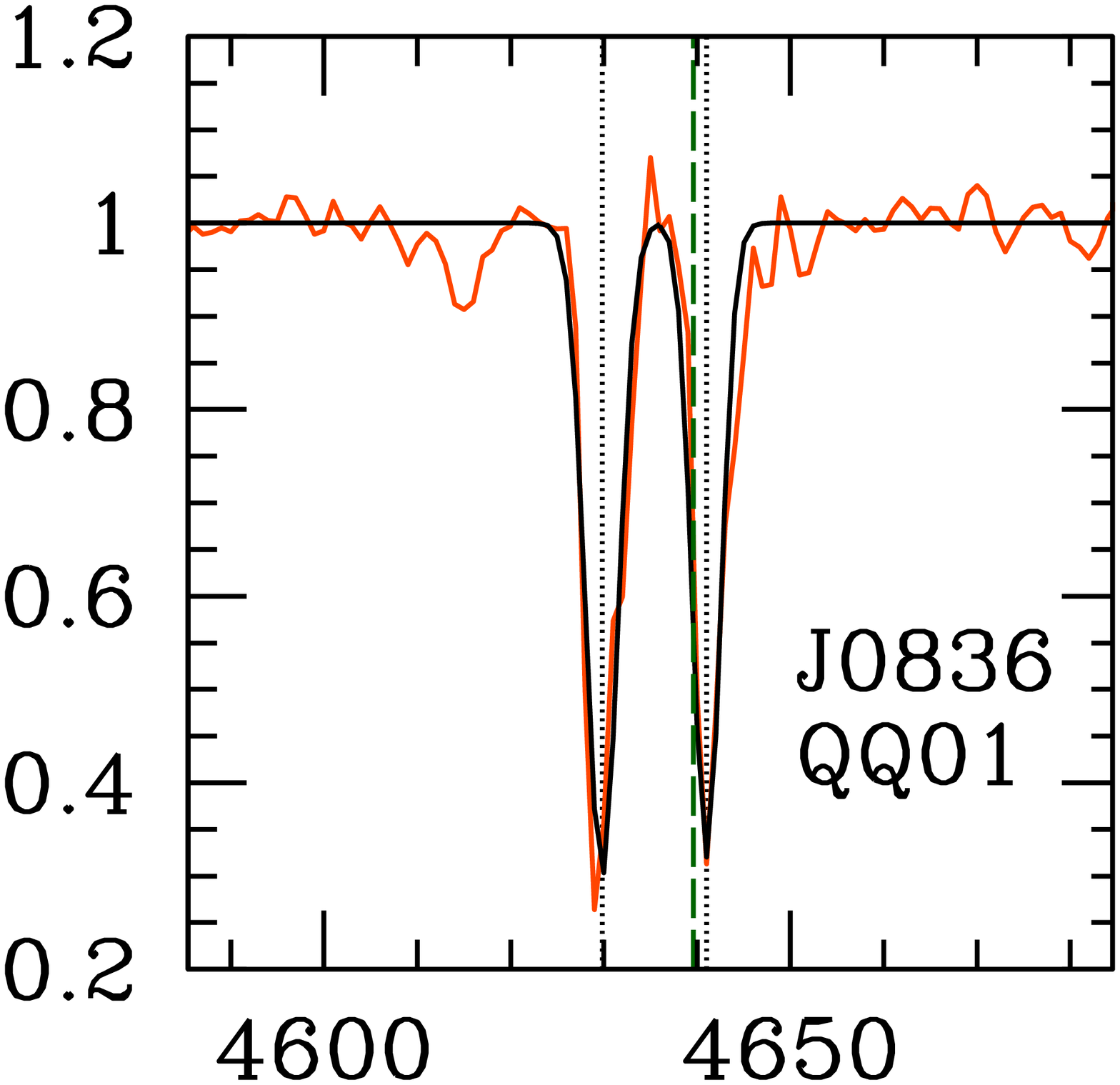}
\includegraphics[width=0.44\columnwidth]{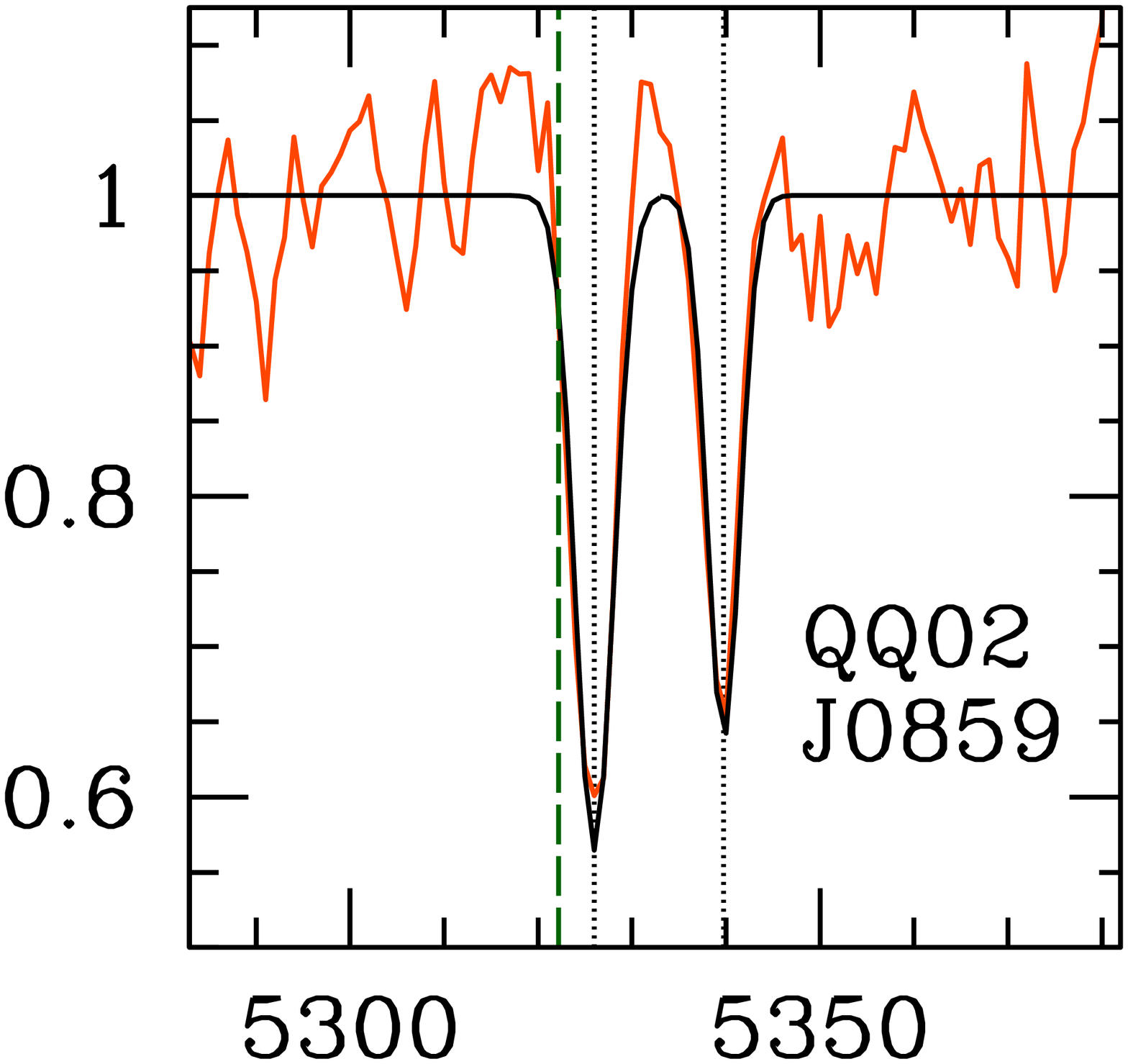}
%\hspace{0.2cm}
\includegraphics[width=0.44\columnwidth]{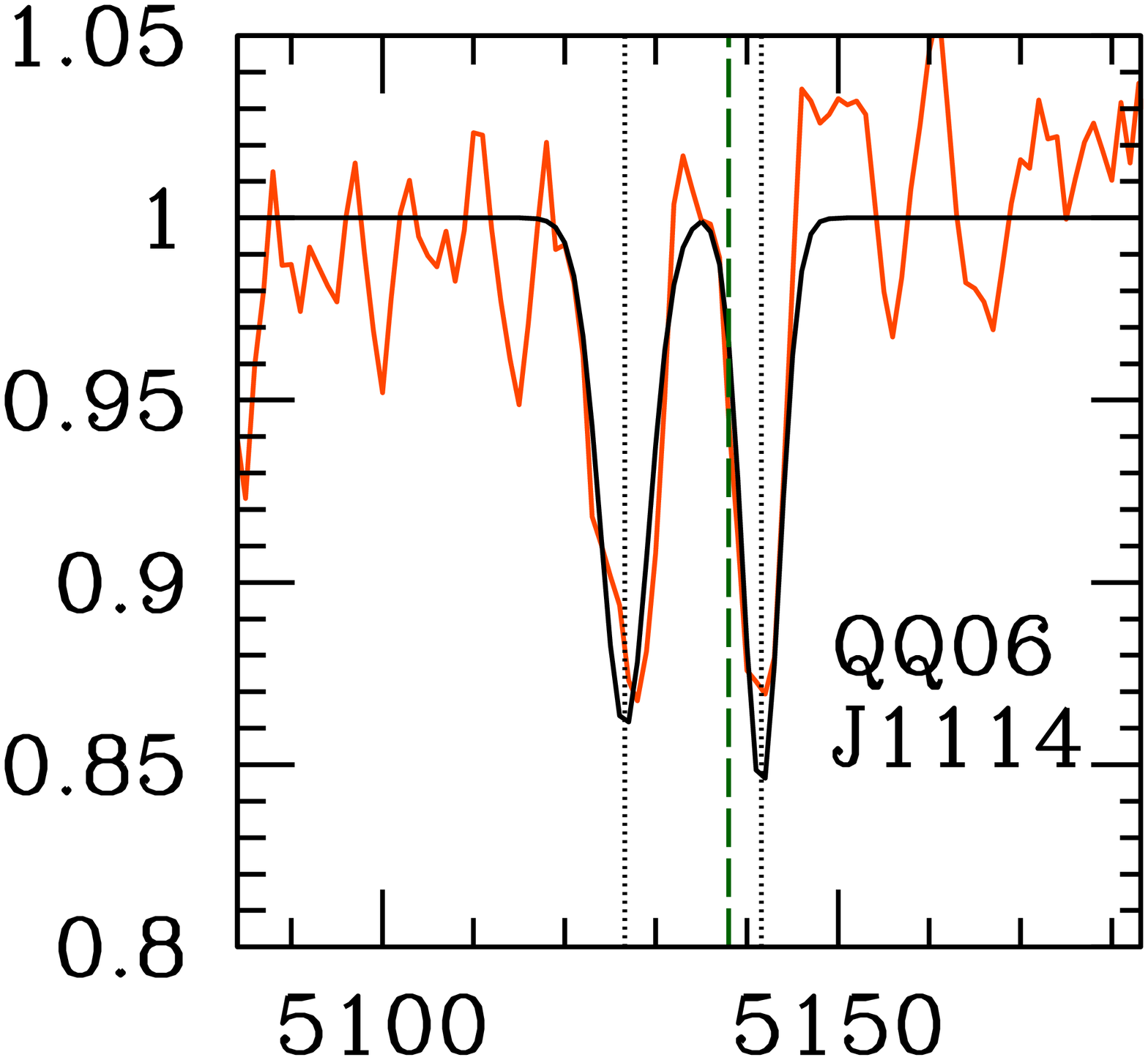}
%\hspace{0.2cm}
\includegraphics[width=0.44\columnwidth]{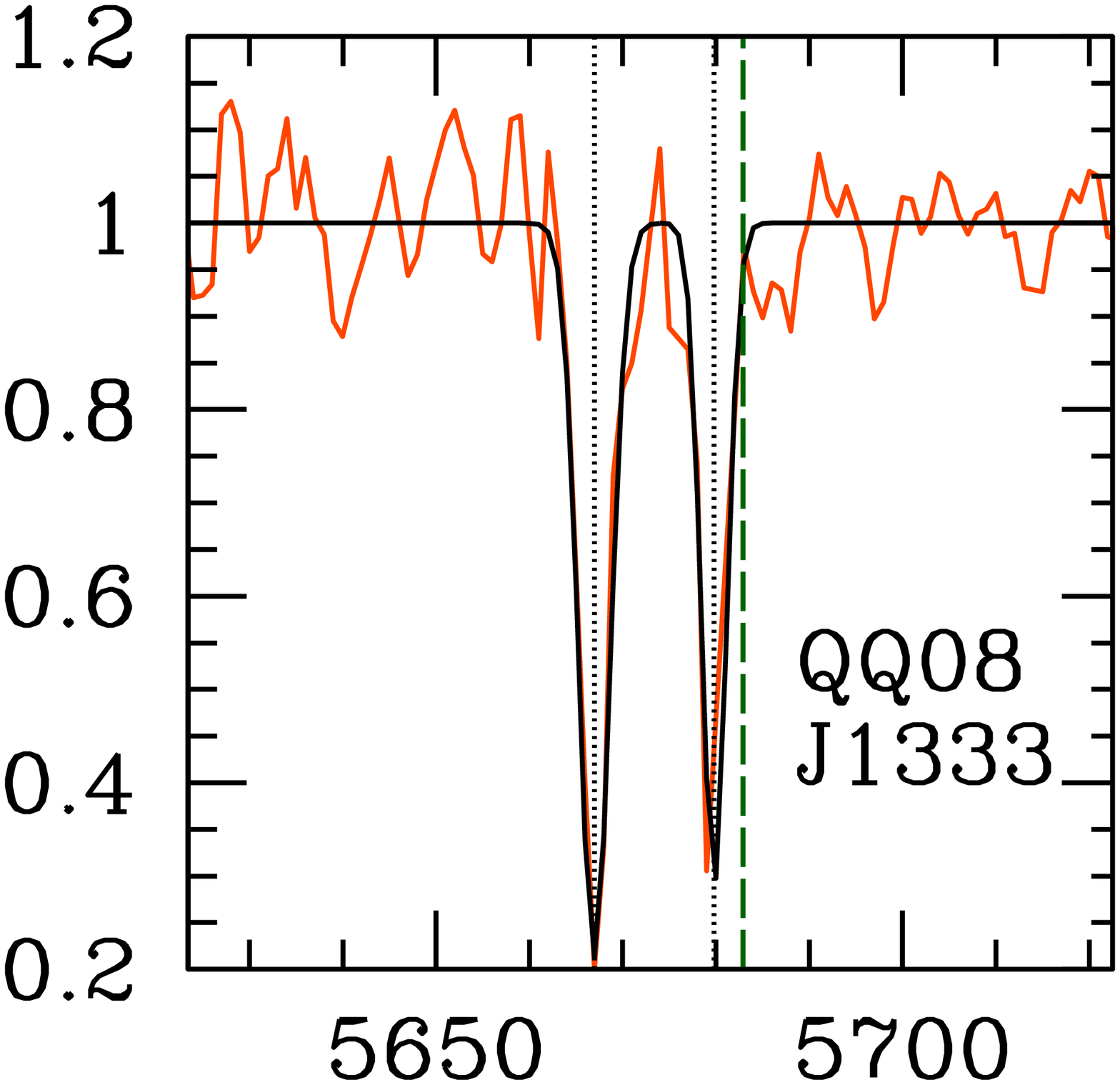}
\includegraphics[width=0.44\columnwidth]{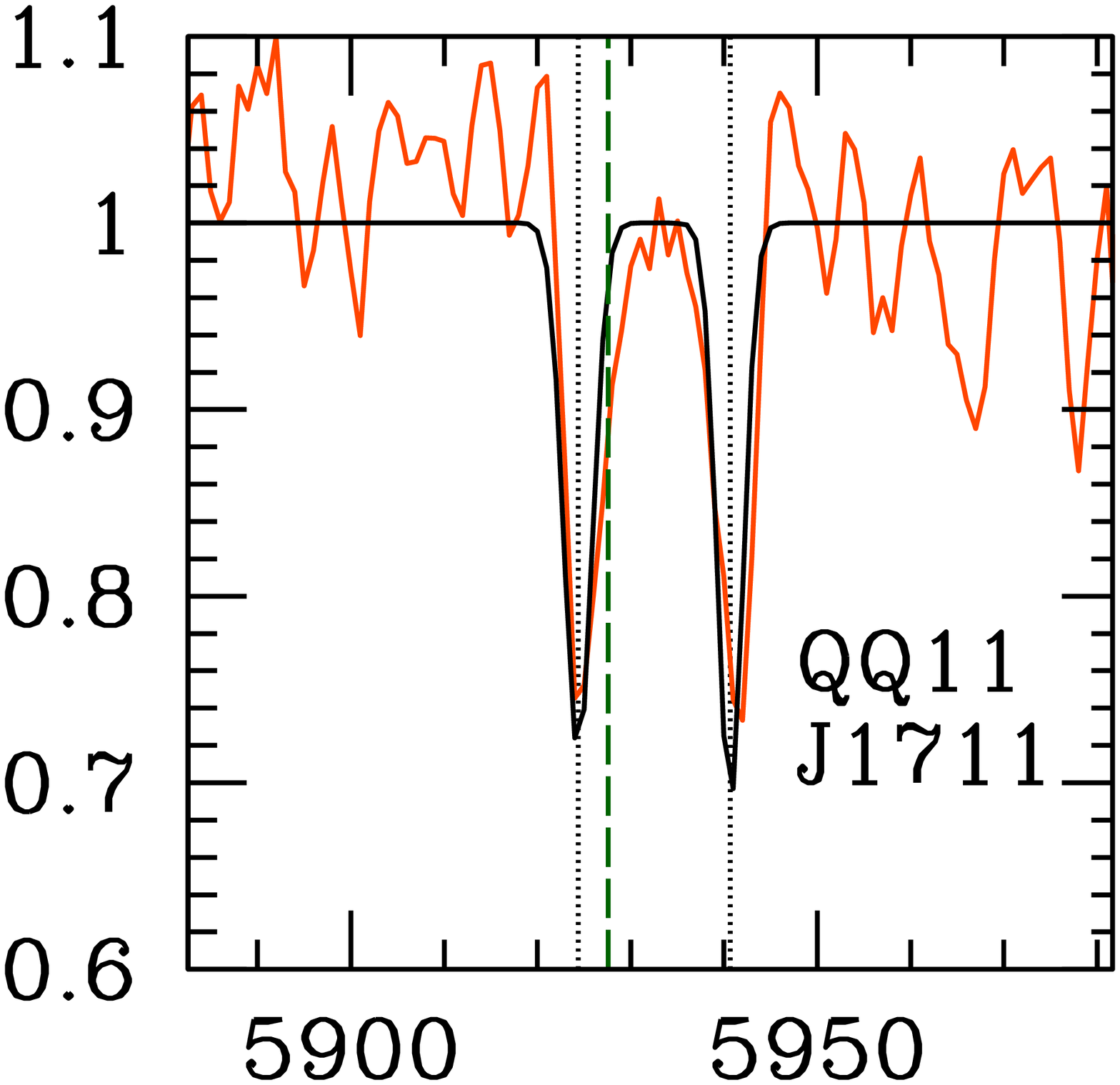}
\includegraphics[width=0.44\columnwidth]{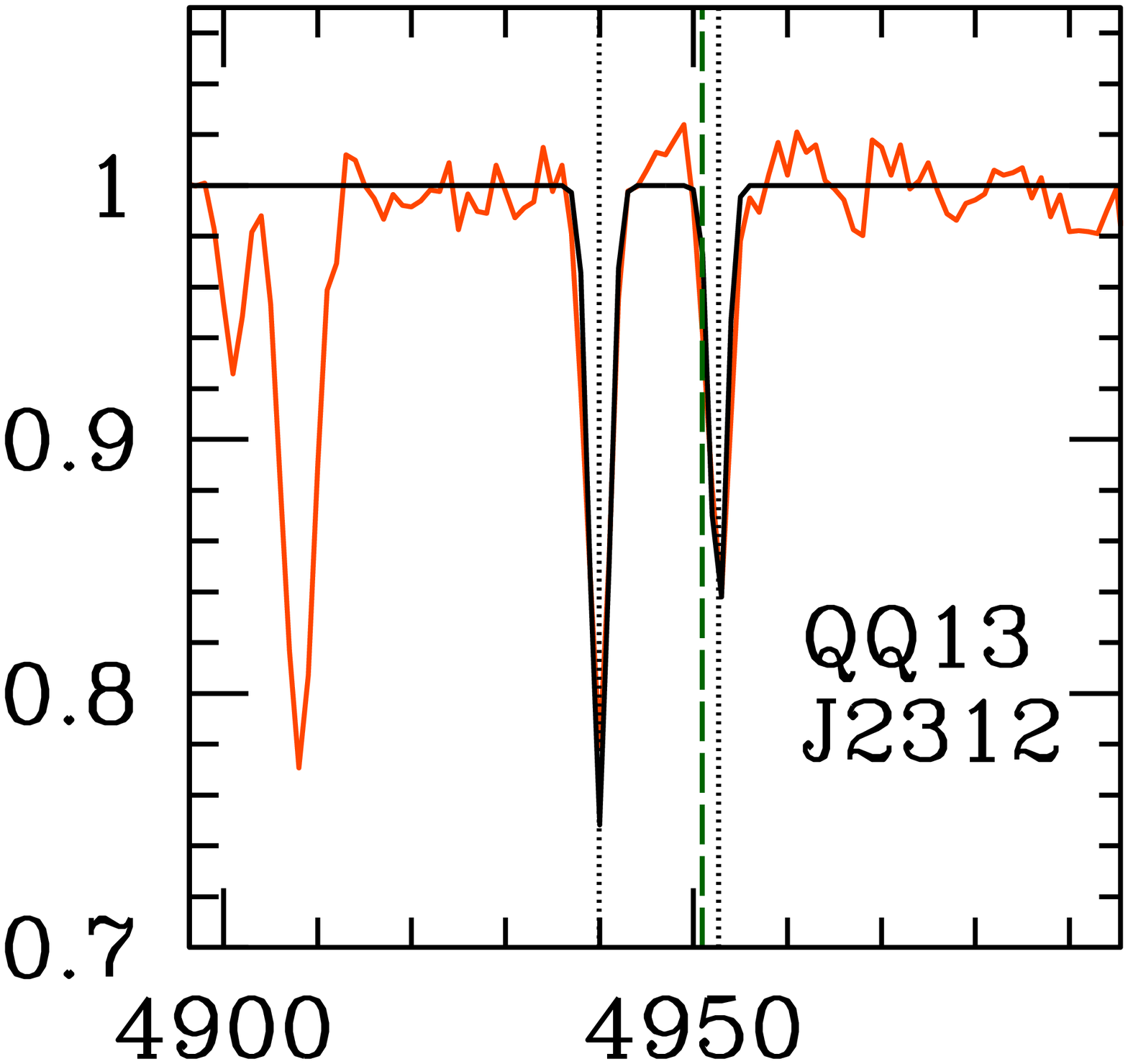}
\includegraphics[width=0.44\columnwidth]{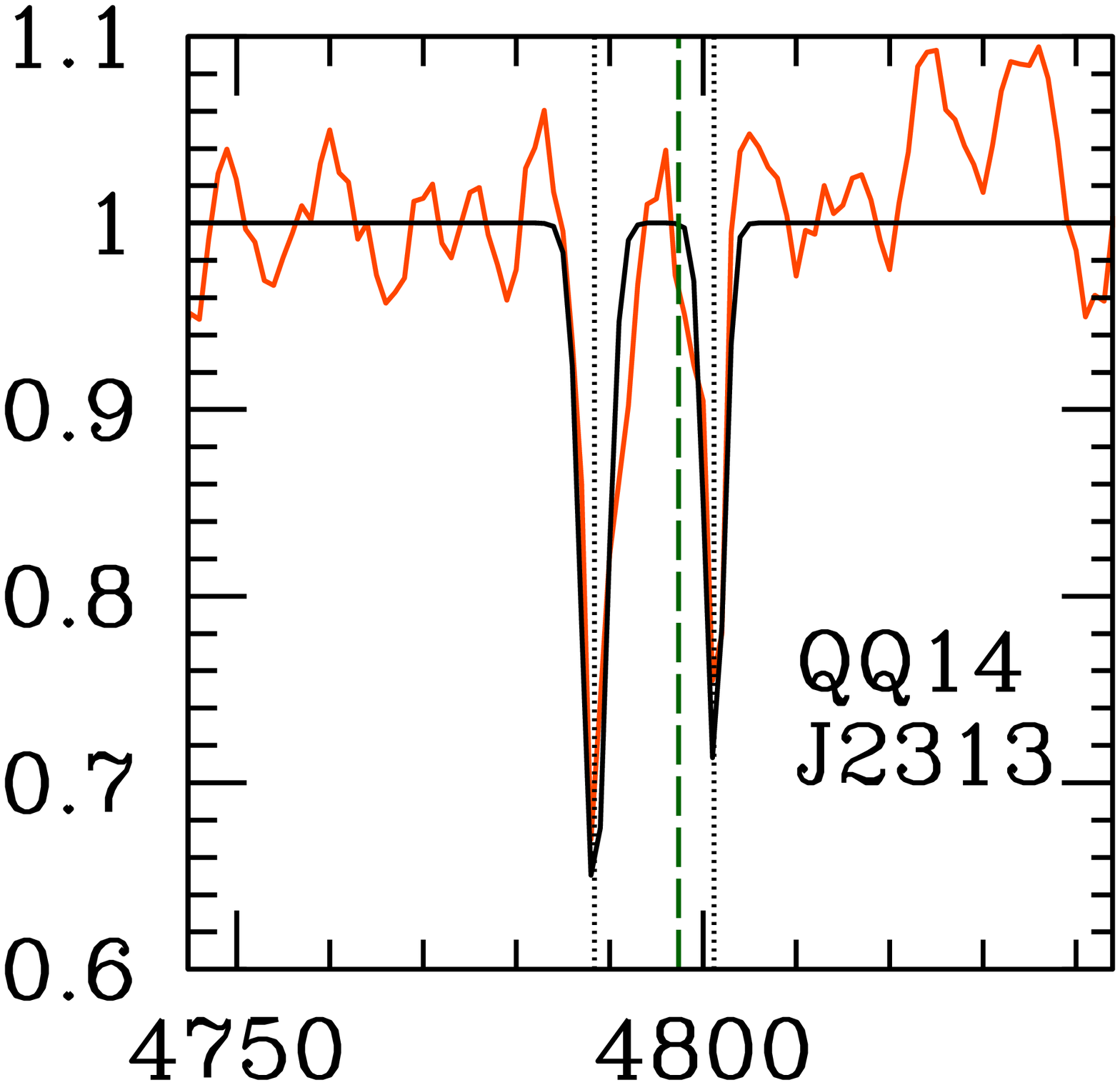}

\label{zoom}
\end{figure*}

We detect transverse MgII absorptions in 7 pairs and close-ups of 
the absorptions with gaussian fits are reported in Fig. \ref{zoom}.
 The covering fraction $f_C$(0.3 \AA), i.e. the fraction of cases where  \mgII 
 absorptions are apparent 
with a sensitivity threshold corresponding to a rest-frame equivalent width
\ewrest($\lambda$2796) $\gtrsim$ 0.3 \AA \ ,  
 in our new sample is  50\%.

\begin{figure}
\centering
\includegraphics[width=0.9\columnwidth]{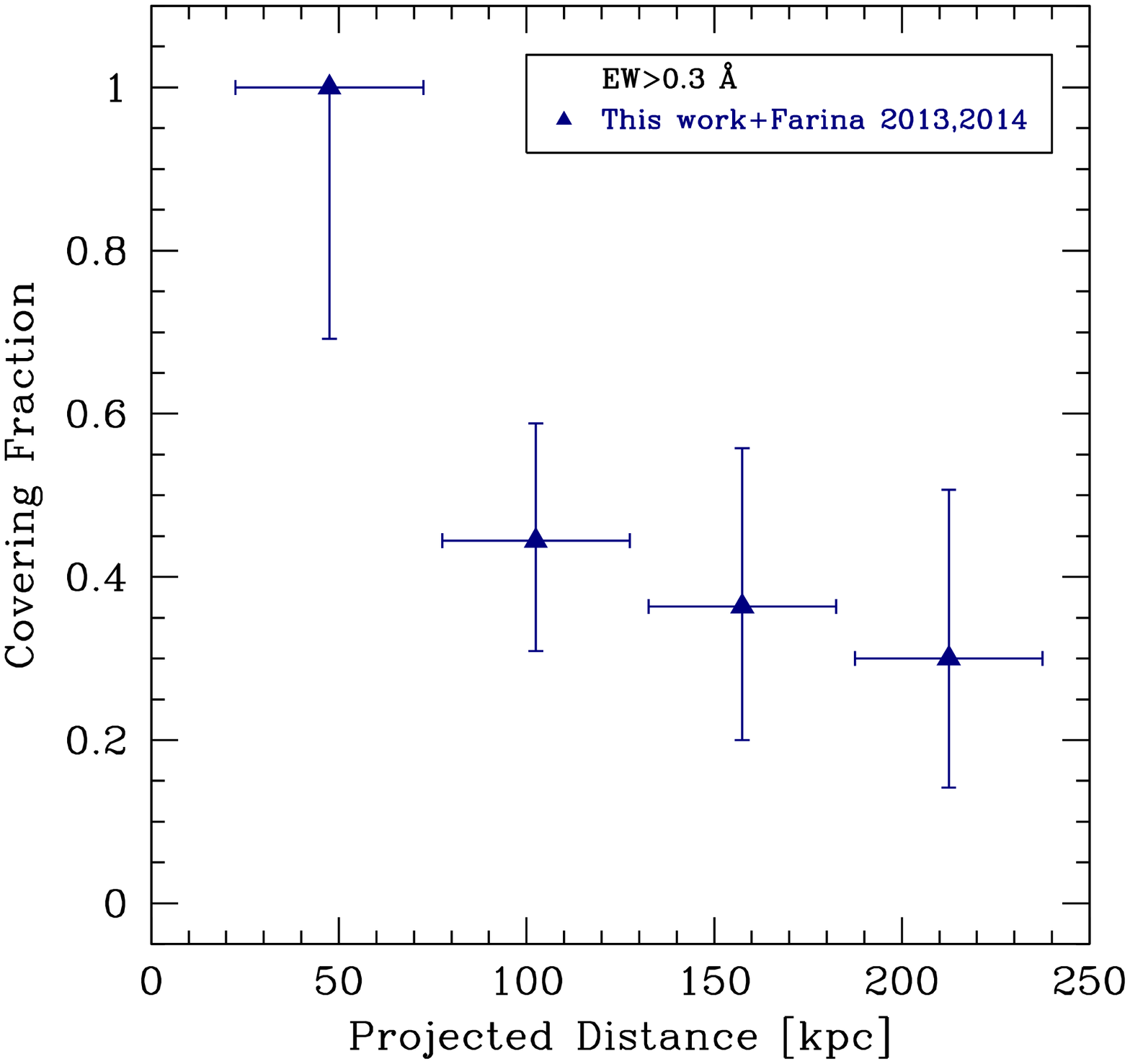}
%\vspace{-2cm}
\caption{
Covering fraction profile for transverse  absorption systems in the QSO$_B$ spectrum  associated with 
the QSO$_F$ with EW(2796) $>$ 0.3 \AA \ , plotted as a function of projected distance between the two QSOs. 
Triangles refer to absorption systems investigated  in  our works, this  and Farina et al. 2013, 2014.
Horizontal bars represent the impact parameter of projected distance, which span from 20 to 240 kpc in 55 kpc wide intervals.
 In each bin the 1$\sigma$ binomial confidence intervals is reported as vertical error bar \citep{Gehrels1986}.  
 }
\label{fc}
\end{figure}

We explore  the extent and the properties of the MgII low-density absorbing gas 
located  in the CGM environment of quasars  using the total sample of 44 projected pairs 
located at 0.6 $\lesssim$ z$_F$ $\lesssim$ 2.2 and investigated by our team in this work and in \FF. 
Over the whole range of projected distances ($\sim$ 30 kpc $<$ R$_\bot$ $<$ $\sim$ 250 kpc) we find  
$f_C$(0.3 \AA)=0.45 and $f_C$(0.6 \AA)= 0.25. 
In Fig. \ref{fc}, we show the MgII covering fraction profile  
of  transverse  absorption systems  with EW$_{rest}$(2796) $>$ 0.3 \AA \  against the  impact parameter,
where binomial 1$\sigma$   confidence intervals  are taken as uncertainties  \citep{Gehrels1986}.
The covering fraction is $fc$(0.3 \AA) = 1.00$^{0.00}_{0.69}$
in the first bin (20 kpc $<$ R$_\bot$ $< $75 kpc) and  decreases
with the impact parameter. Note however that we observe the presence of three 
absorbers over 200 kpc, and this yields   $fc$(0.3 \AA) = 0.30$^{0.51}_{0.14}$
in the range  185  kpc $<$ R$_\bot$ $<$ 240 kpc.

\section{Line of sight absorptions }

We now focus on LOS absorptions, which are absorption features superposed to the \mgII emission line of the \qsof. 
For attributing the absorption to the CGM cloud \cite{Farina2014} considered two possible velocity 
difference thresholds 1000 km/s, or  following  \cite{Sharma2013}, 5000 km/s. 

\begin{figure}
\centering
\includegraphics[width=0.49\columnwidth]{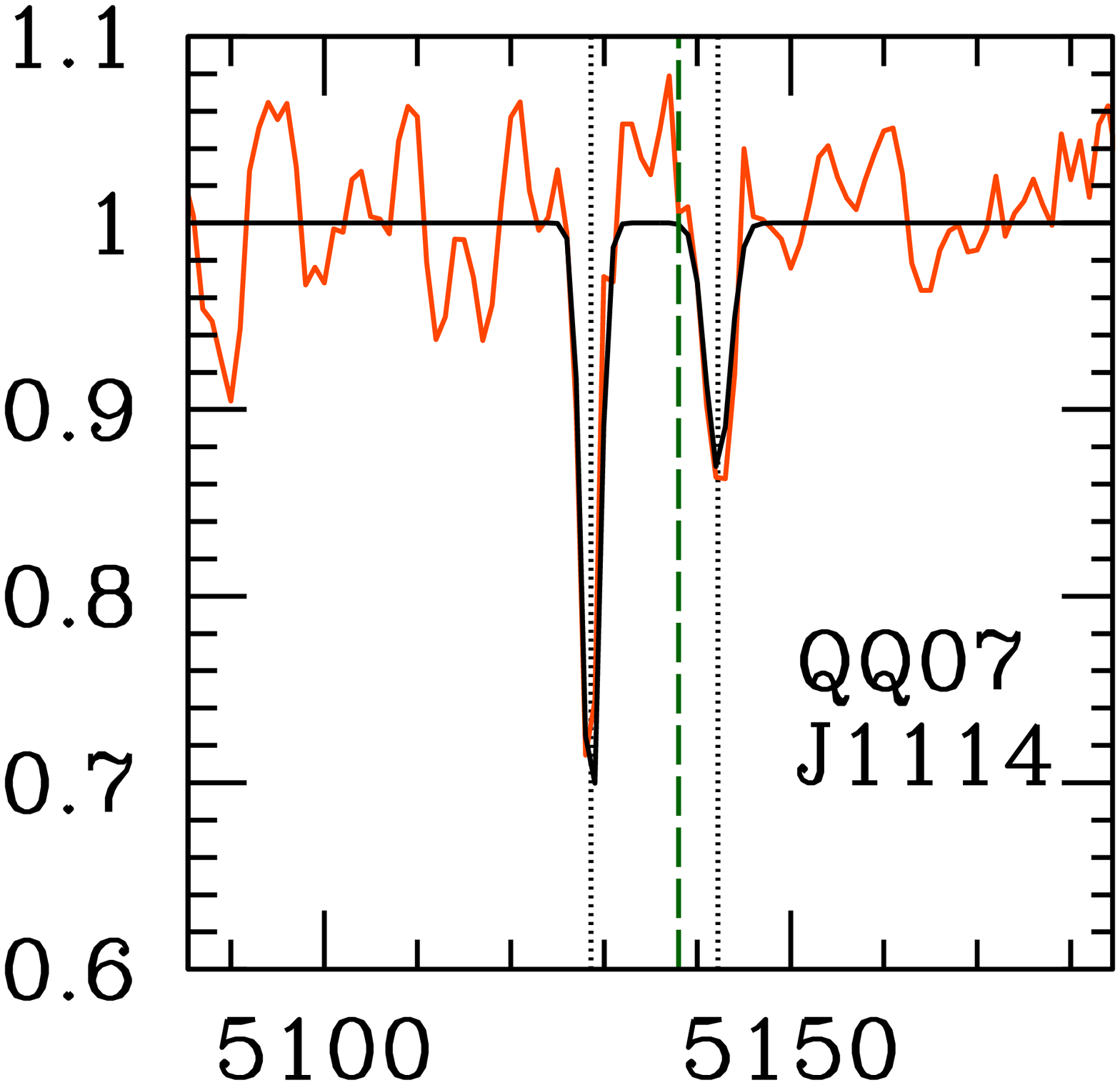}
\includegraphics[width=0.49\columnwidth]{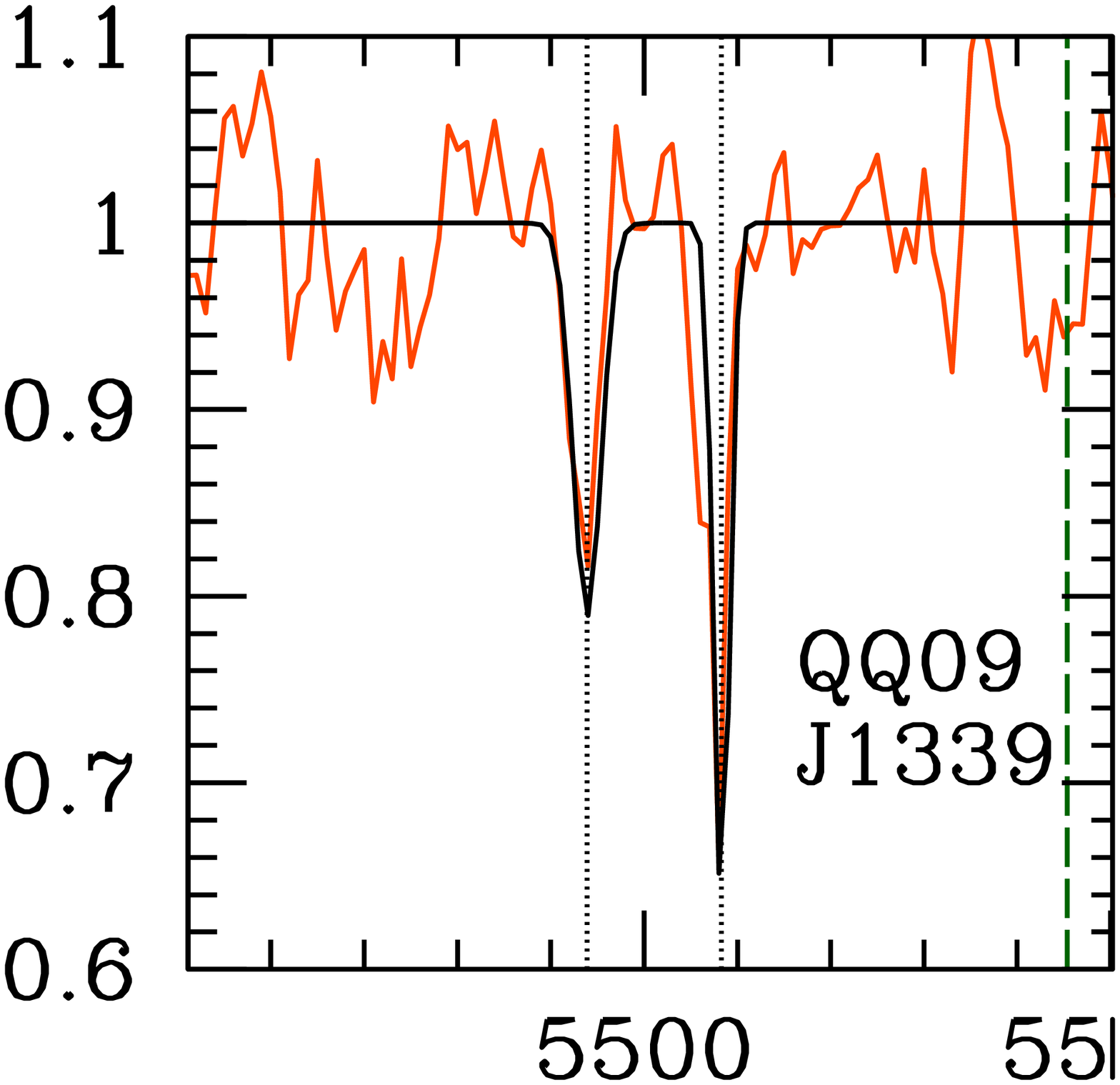} 
\caption{
Close-ups of the normalized \qsof \  spectra (red lines)    presenting line-of-sight absorption systems  associated to the 
\mgII  emission line of the \qsof \ itself. The spectra are normalised along the MgII emission line profile.
Gaussian fits are  performed on the absorptions lines and drawn as solid black lines. Dotted black lines 
indicate the positions of  spectral absorption peaks, while the green dashed line marks the position of  the \mgII emission line
in the  \qsof \ spectrum.
}
\label{zoom_los}
\end{figure}

Taking the former value, we find that within our sensitivity 
one LOS absorption (QQ07)  is detected, see Figs.\ref{spec} and \ref{zoom_los},  and Table \ref{los}. 
With a velocity limit of 5000 km/s, we find another case of LOS absorption,
reported in Figs.\ref{spec}, \ref{zoom_los} and Table \ref{los}, with $\Delta v_{rest} \sim$ 2000 km/s, 
to be compared to the one case of \cite{Farina2014}.  

For the paucity of LOS with respect to transverse absorptions \cite{Farina2014} suggest as a possible explanation 
that the LOS absorbing clouds are heated at a temperature such that \mgII is practically absent, but this is not the case in the transverse direction. 
 This implies an anisotropy of the continuum emission. This picture was originally   proposed by \cite{Prochaska2013} for the Ly$\alpha$ LOS absorptions. 
 Our current dataset is still too small  to further elaborate on  this  proposal.
  Since the distribution of the gas in the external regions is patchy a sound statistical approach is required.  
  A  larger dataset of high quality and homogeneous spectra of quasar projected pairs would be able to draw firm conclusions on this intriguing scenario.

\begin{figure*}[h!]
\caption{Normalized QSOs spectra.
The spectra of the  \qsof \  and the \qsob \  are displayed at
 the top and at the bottom panel of each frame, respectively.
The main emission lines are labeled and unidentified by dashed black lines. 
Detected  MgII absorption features in the \qsob \ spectrum associated
with the \qsof \ are  evidenced by a yellow bar,  the redshift of the absorbers are reported in green. 
The positions of line-of-sight (LOS)
absorbers in the foreground QSO  spectra are singled out by magenta triangles, see also Figure \ref{los}.
 Red dotted lines mark absorption  systems due to intervening clouds or transverse absorption features of CIV. 
}\label{spec}
%\vspace{0.5cm}
\includegraphics[trim={2.cm 1.5cm 0.2cm 2.5cm},clip=true,width=0.49\textwidth]{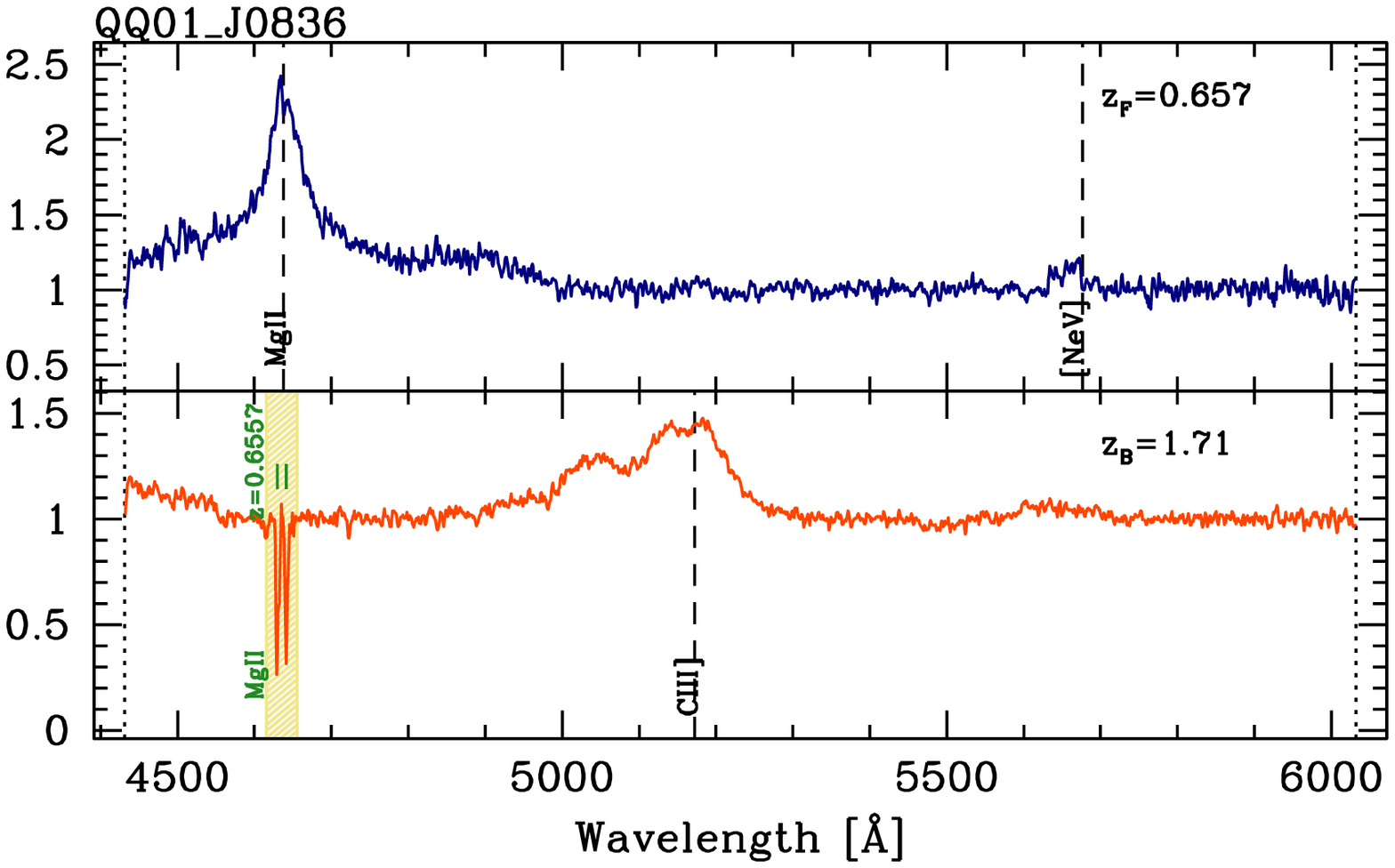}
\includegraphics[trim={2.cm 1.5cm 0.2cm 2.5cm},clip=true,width=0.49\textwidth]{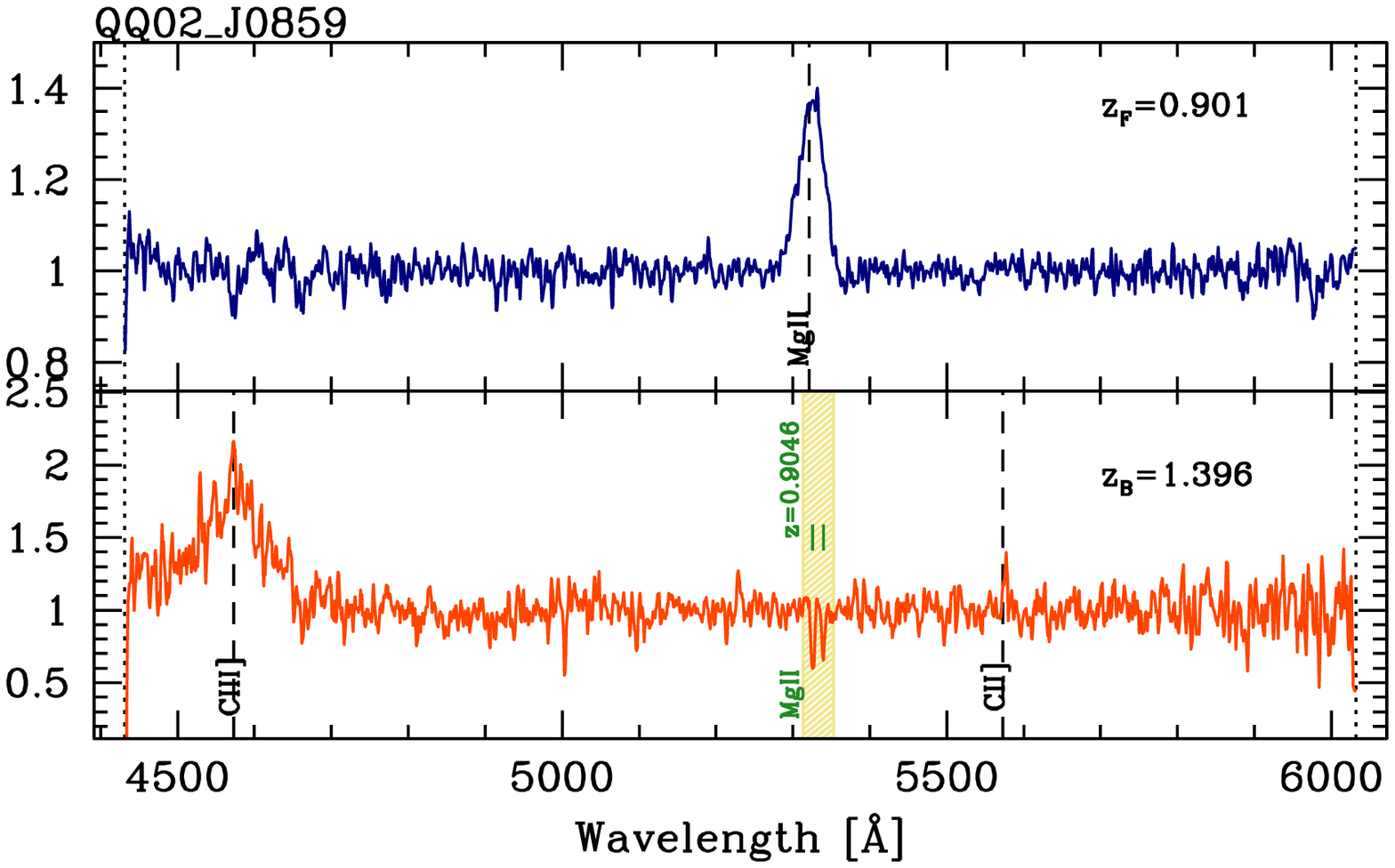}
%\vspace{-3cm}
\includegraphics[trim={2.cm 1.5cm 0.2cm 2.5cm},clip=true,width=0.49\textwidth]{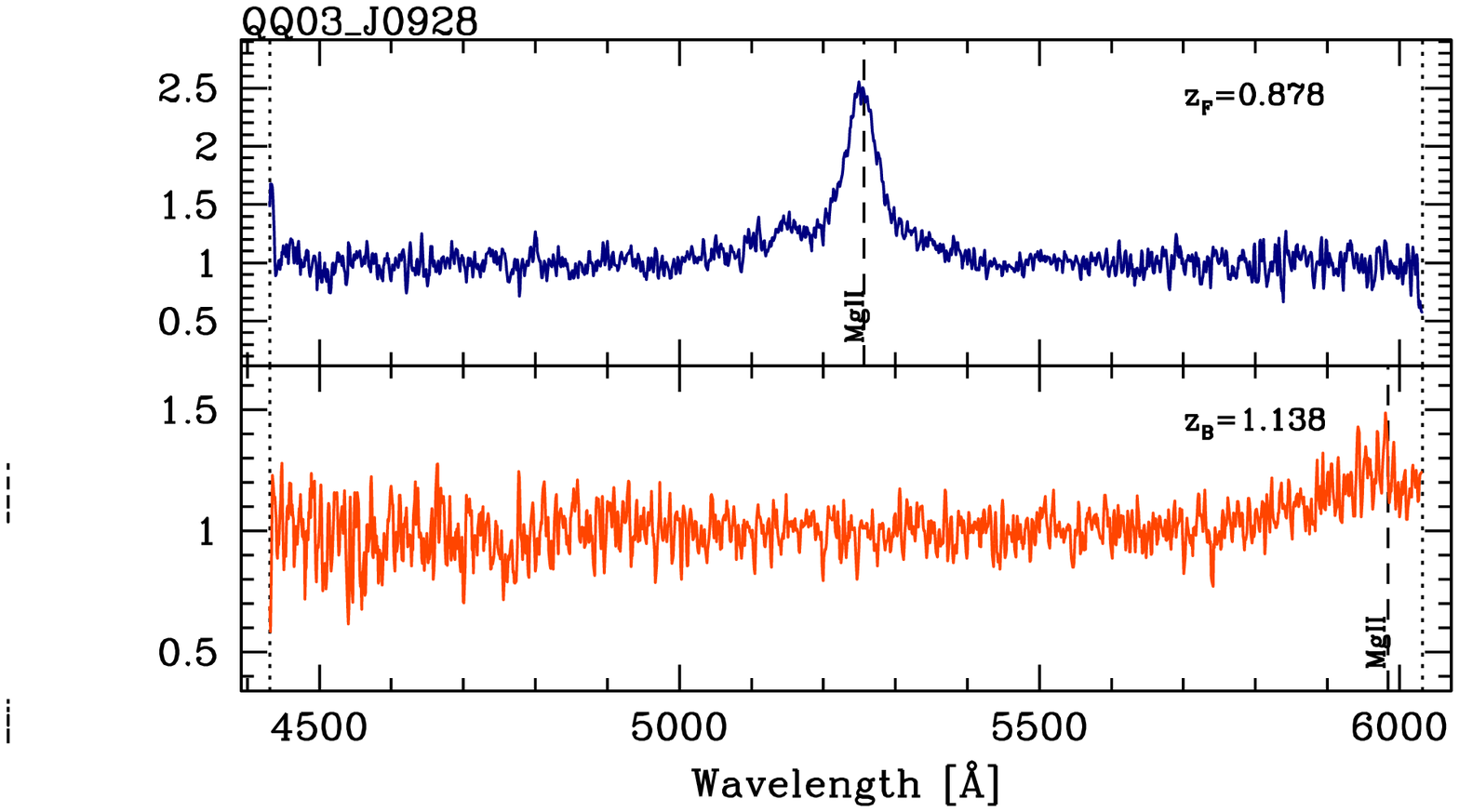}
%\hspace{0.2cm}
\includegraphics[trim={2.cm 1.5cm 0.2cm 2.5cm},clip=true,width=0.49\textwidth]{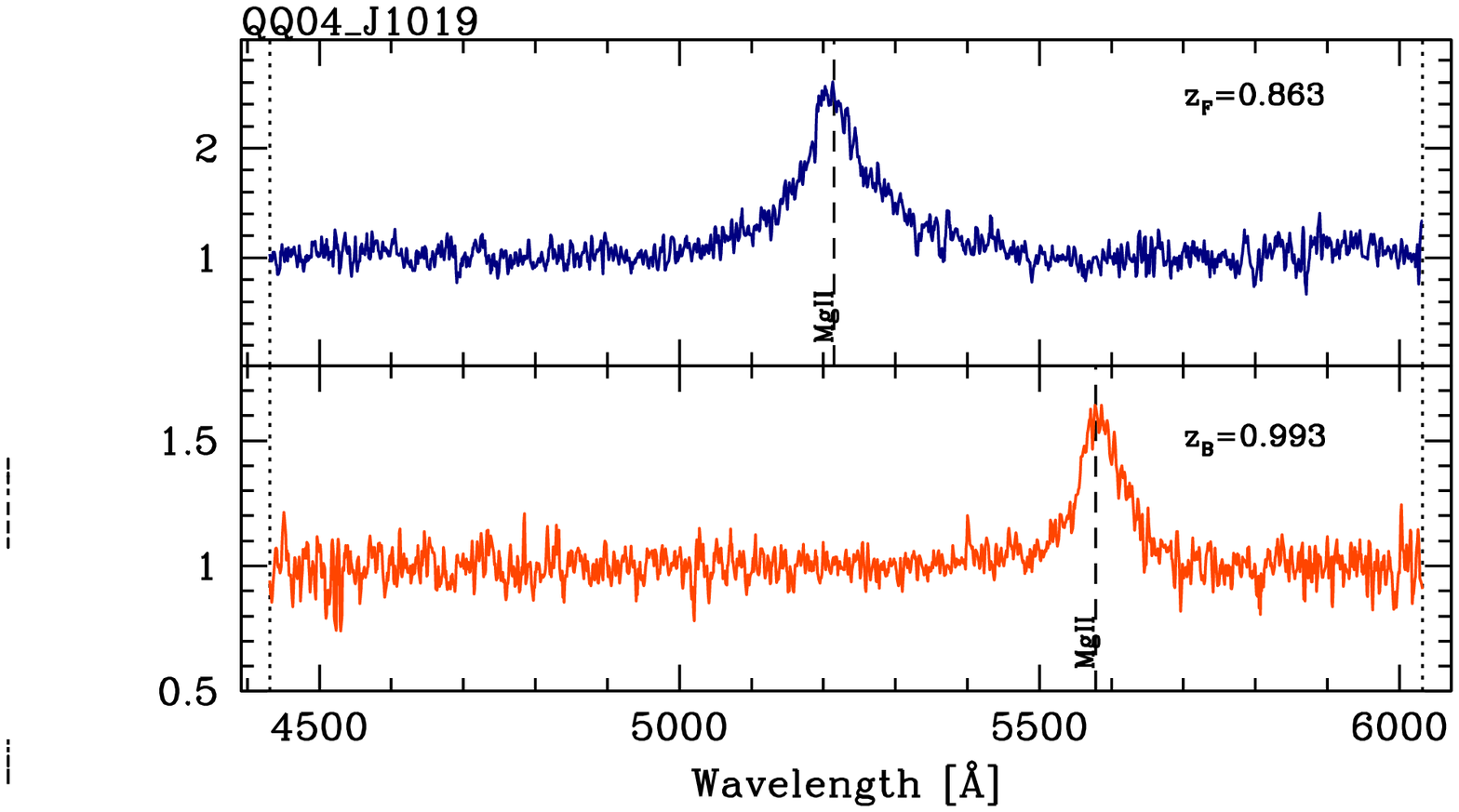}
\end{figure*}

\clearpage

\begin{figure*}
\setcounter{figure}{1}
\caption{...continued...}
\label{spec}
\vspace{-2cm}
\includegraphics[trim={2.cm 1.5cm 0.2cm 2.5cm},clip=true,width=0.49\textwidth]{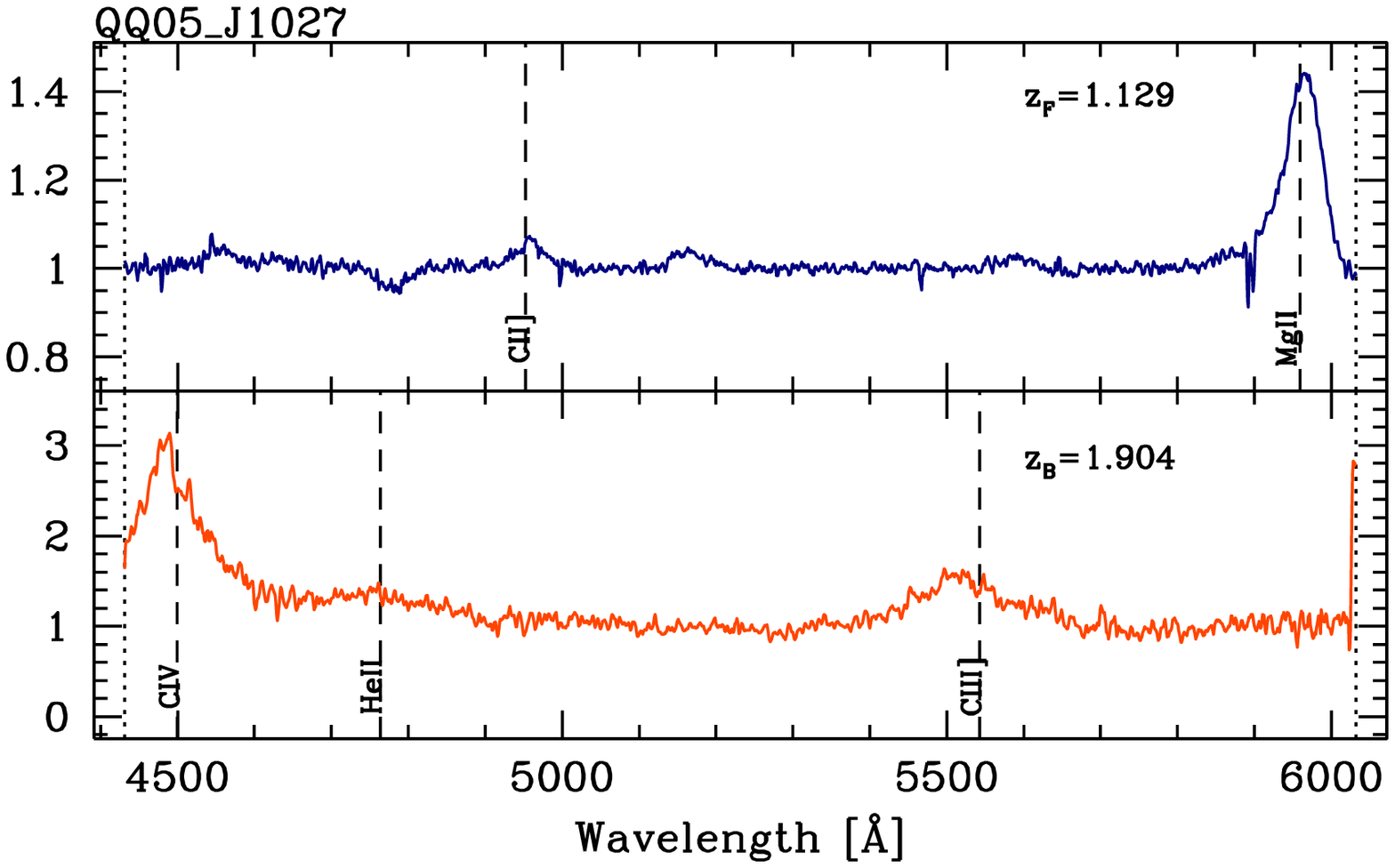}
\includegraphics[trim={2.cm 1.5cm 0.2cm 2.5cm},clip=true,width=0.49\textwidth]{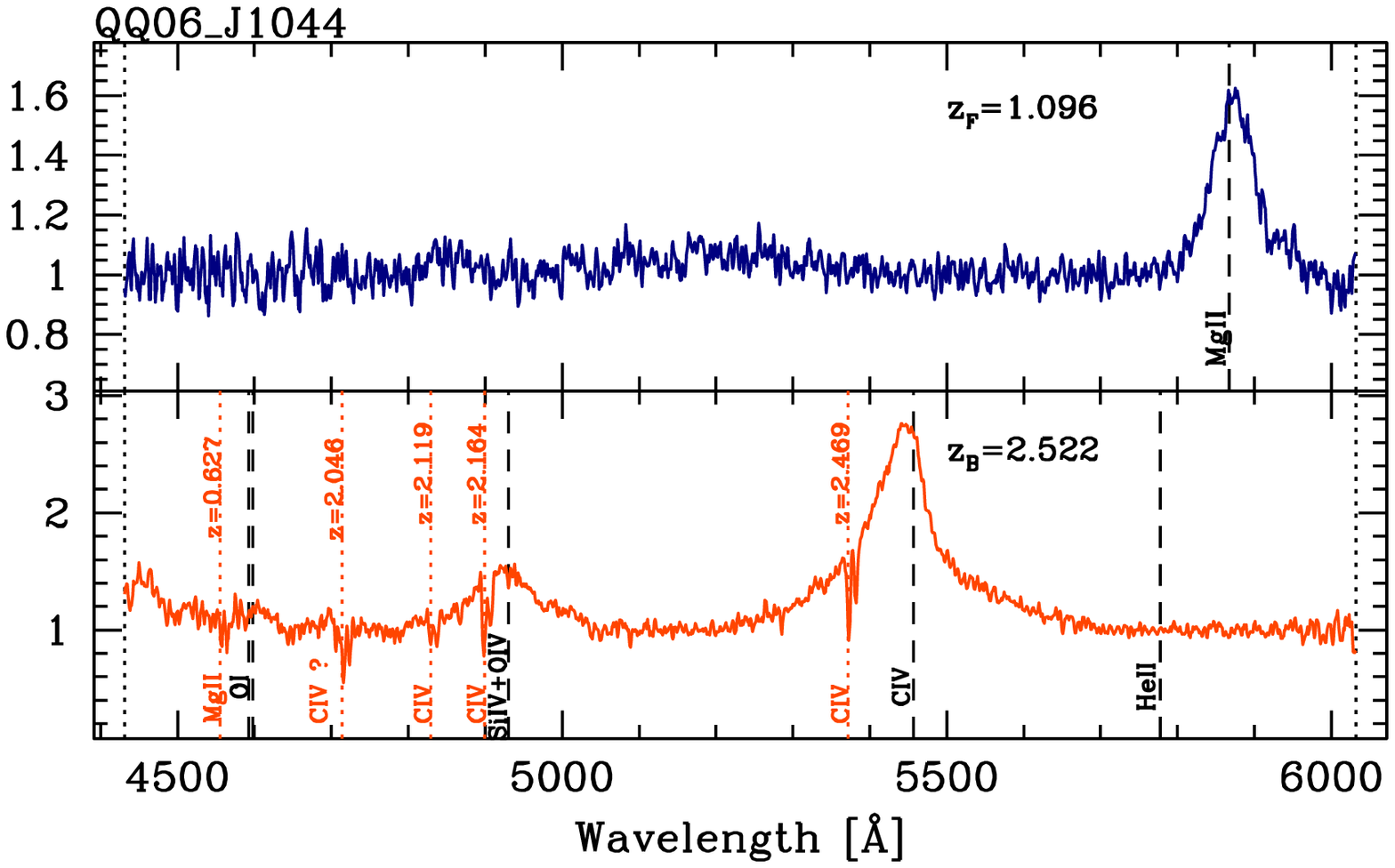}
\includegraphics[trim={2.cm 1.5cm 0.2cm 3cm},clip=true,width=0.49\textwidth]{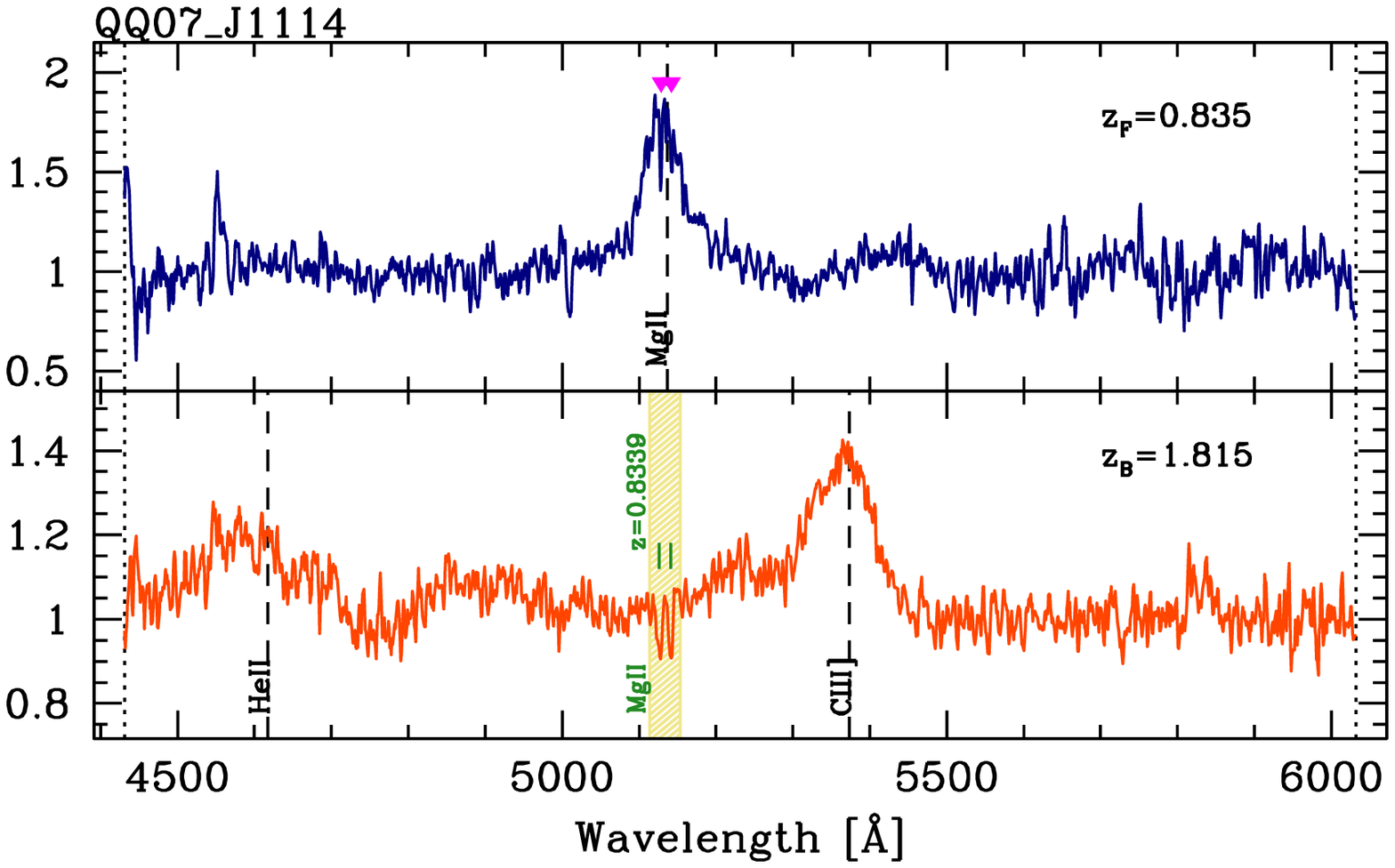}
\hspace{0.2cm}
\includegraphics[trim={2.cm 1.5cm 0.2cm 3cm},clip=true,width=0.49\textwidth]{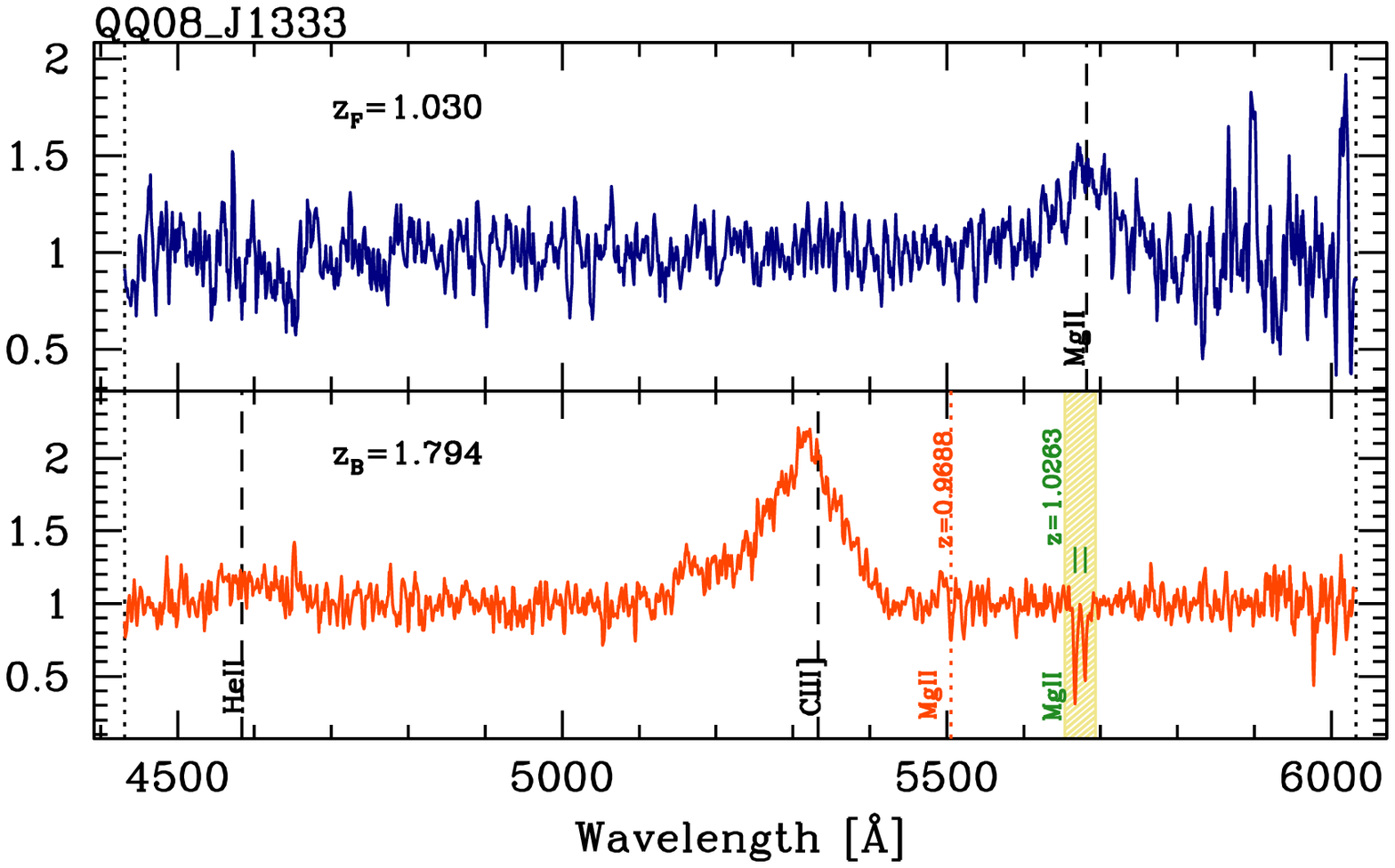}
\includegraphics[trim={2.cm 1.5cm 0.2cm 3cm},clip=true,width=0.49\textwidth]{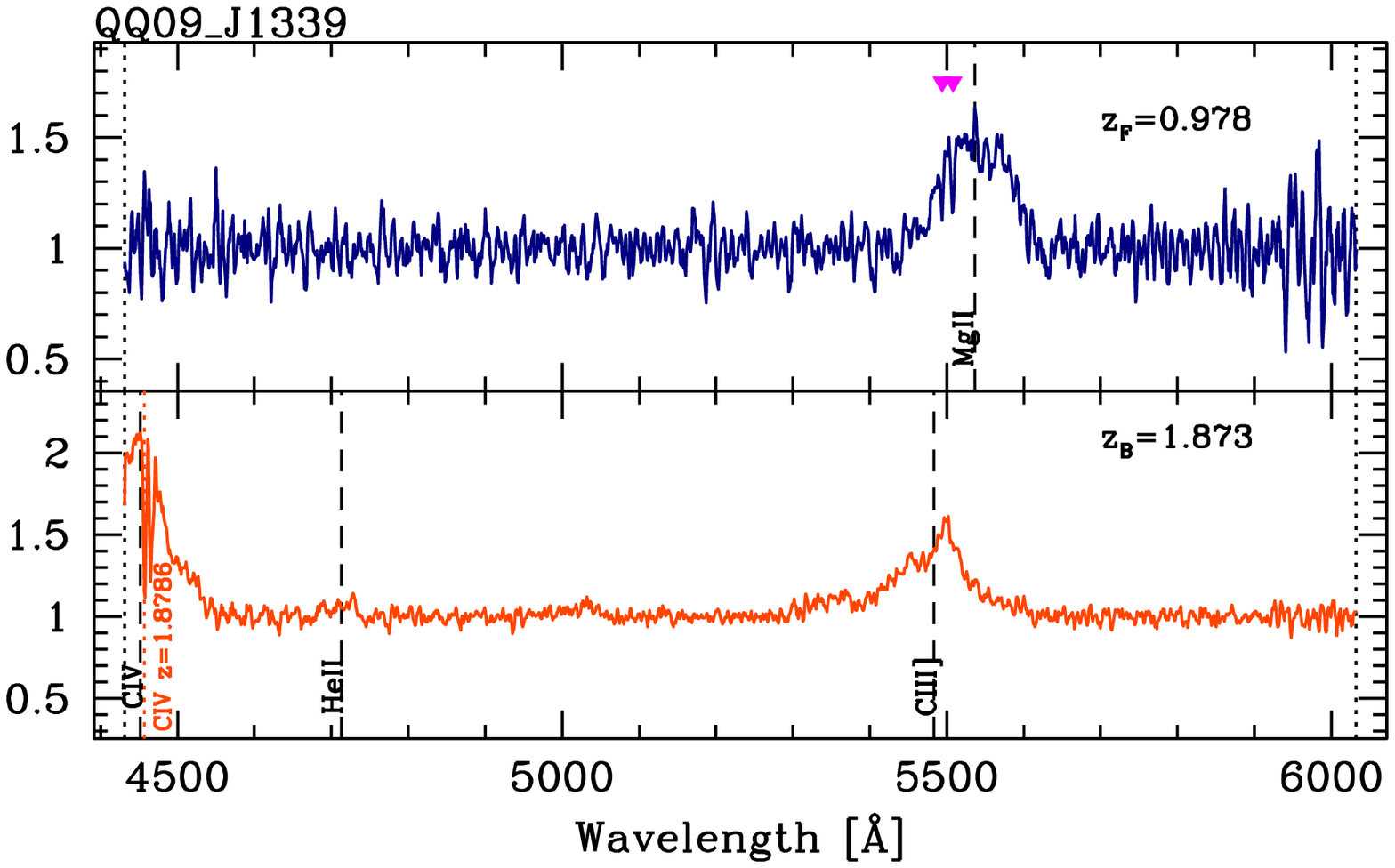}
\hspace{0.2cm}
\includegraphics[trim={2.cm 1.5cm 0.2cm 3cm},clip=true,width=0.49\textwidth]{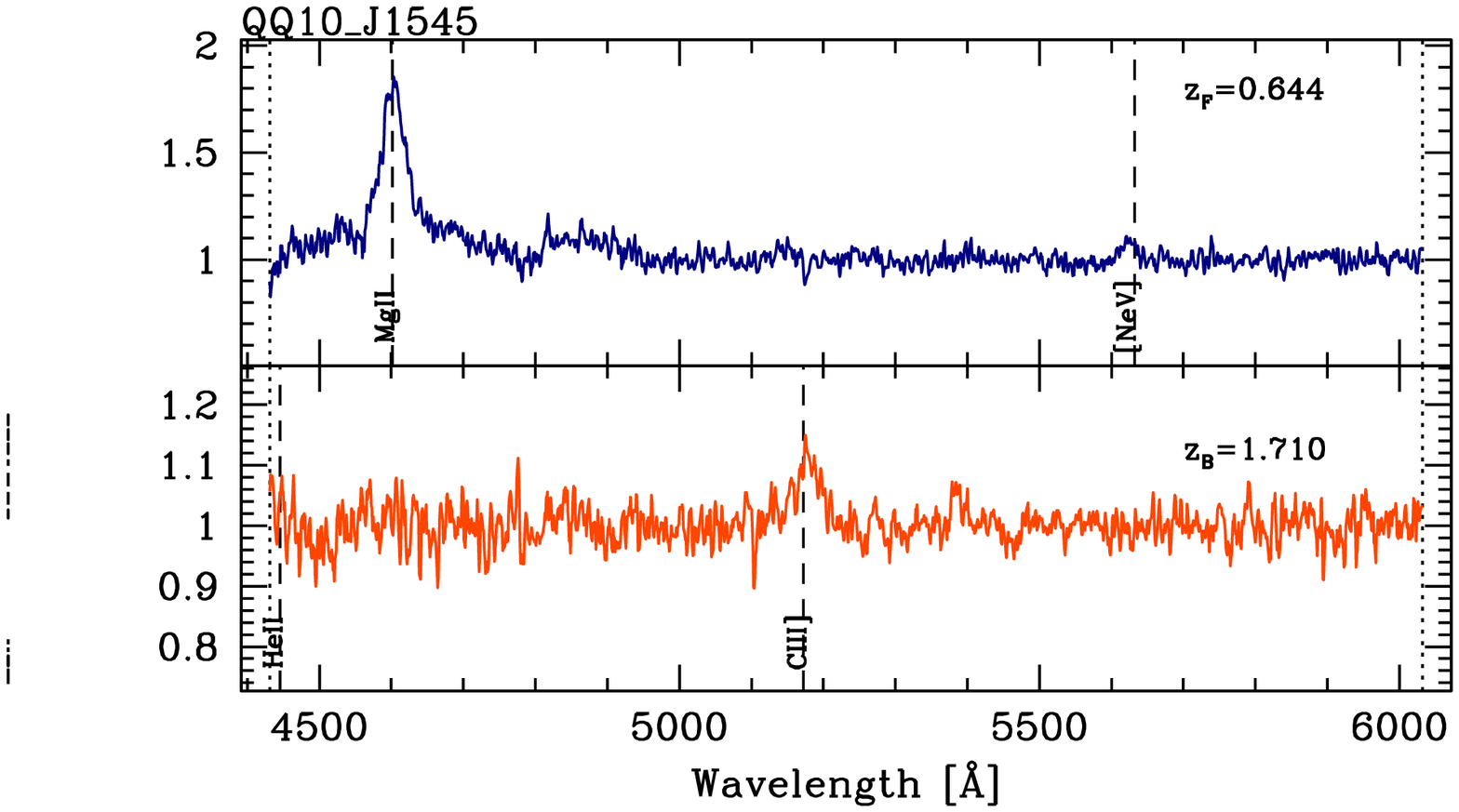}
\end{figure*}

\clearpage 

\begin{figure*}
\vspace{-2cm}
\setcounter{figure}{1}
\caption{...continued.}
\label{spec}
\includegraphics[trim={2.cm 1.5cm 0.2cm 2.5cm},clip=true,width=0.49\textwidth]{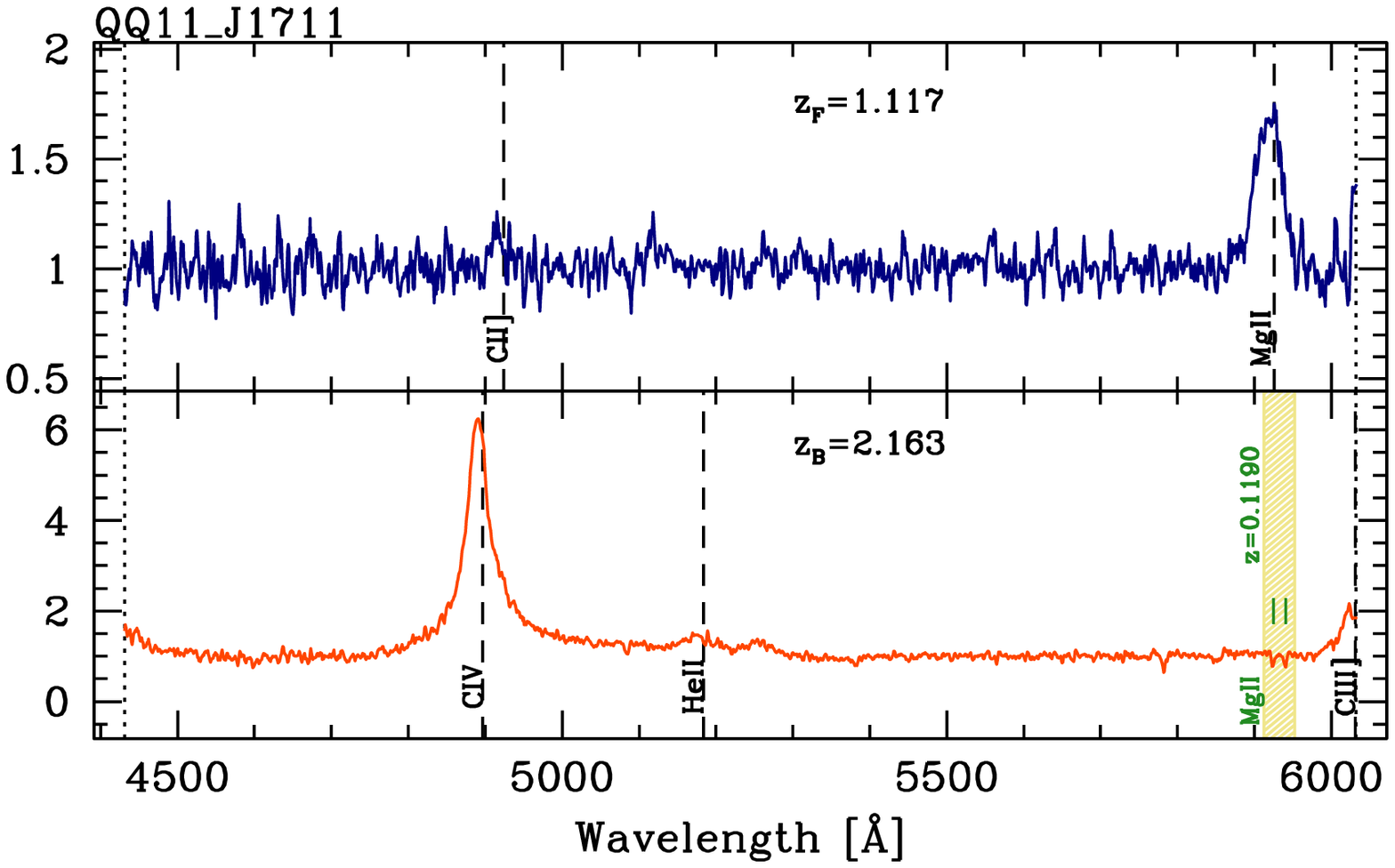}
\includegraphics[trim={2.cm 1.5cm 0.2cm 2.5cm},clip=true,width=0.49\textwidth]{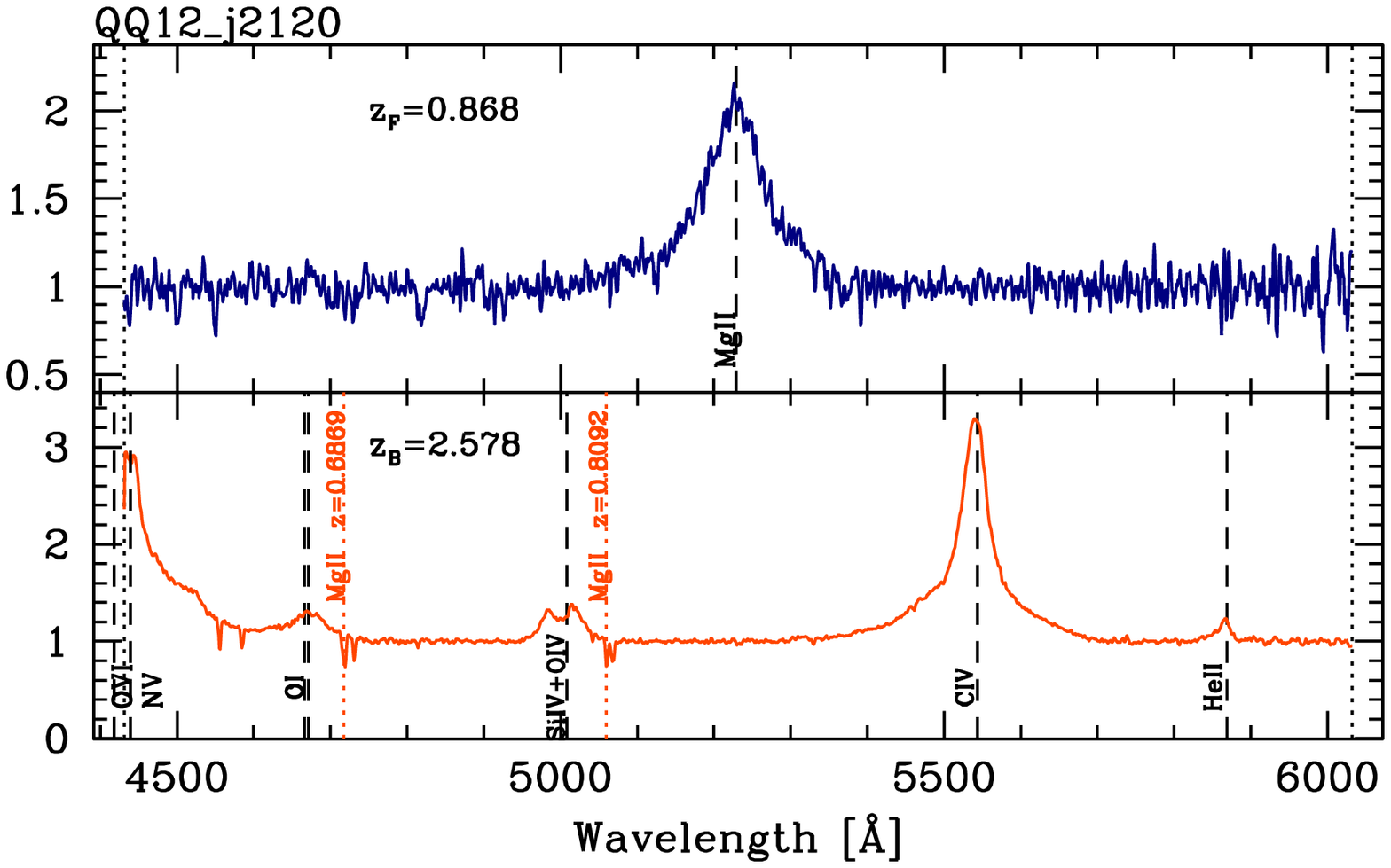}
%\vspace{-2cm}
\includegraphics[trim={2.cm 1.5cm 0.2cm 2.5cm},clip=true,width=0.49\textwidth]{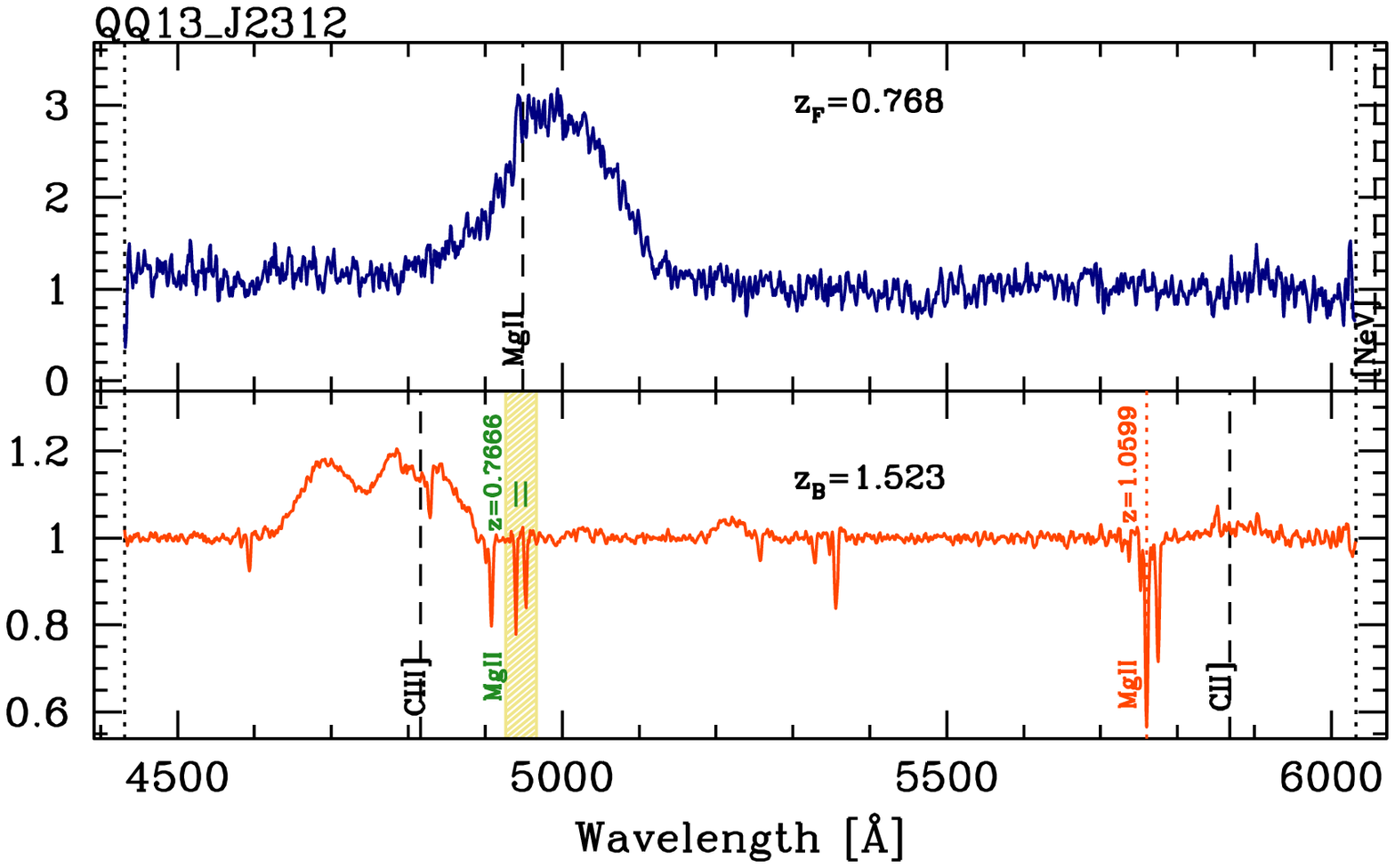}
\hspace{0.2cm}
\includegraphics[trim={2.cm 1.5cm 0.2cm 2.5cm},clip=true,width=0.49\textwidth]{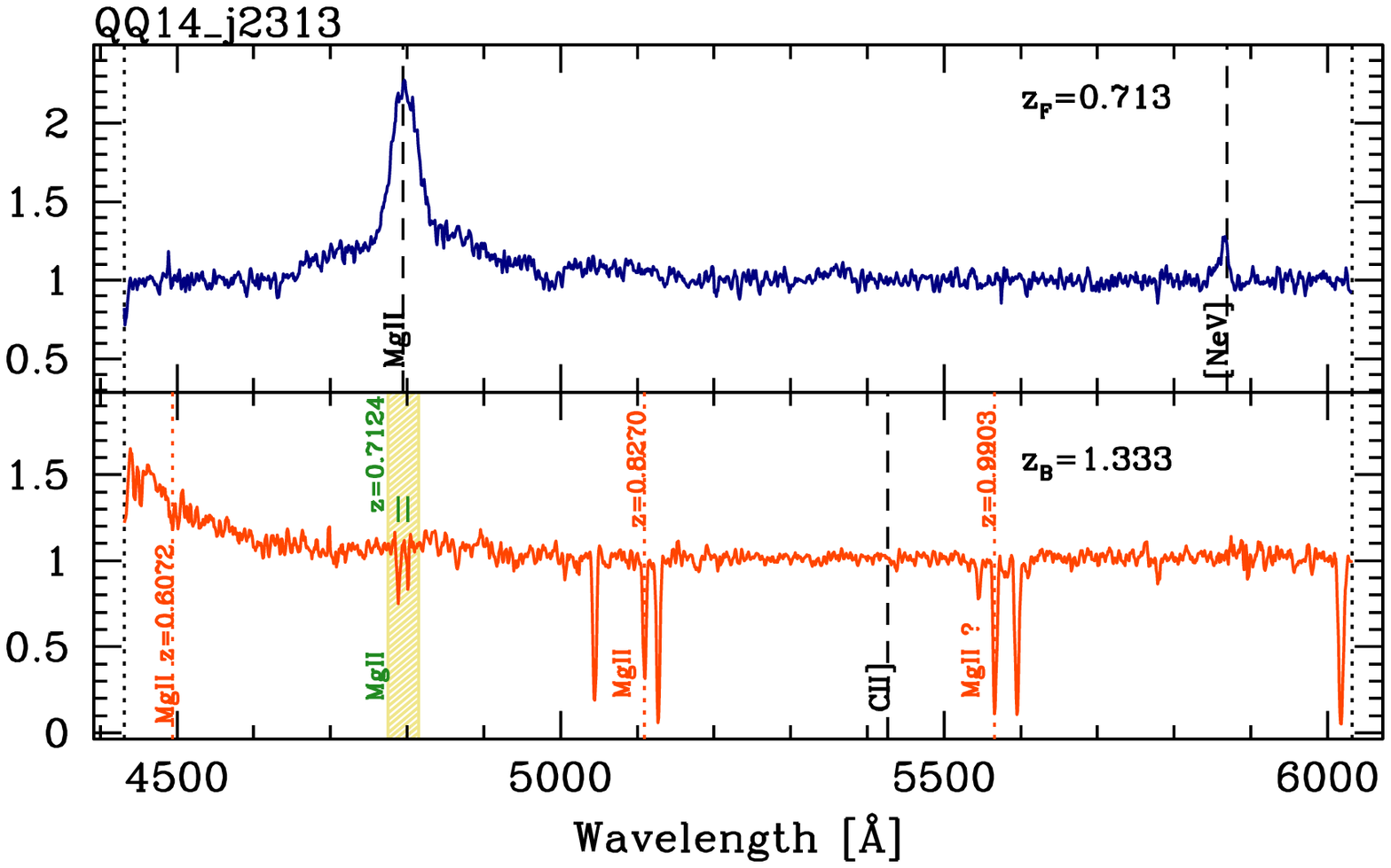}

\label{}
\end{figure*}

\clearpage

%% Acknowledgements
%
% \acknowledgments
% <Acnowledgments text>

%%%%%%%

\end{document}